\documentclass[onecolumn, 12pt, amsmath,amssymb,amsfonts,superscriptaddress,floatfix,showpacs]{revtex4-1}
\pdfoutput=1
\usepackage{graphicx}% Include figure files
\usepackage{dcolumn}% Align table columns on decimal point
\usepackage{keyval}
\usepackage{bm}% bold math
\usepackage{subfigure}
\usepackage{float}
\usepackage{amsmath}
\usepackage[usenames,dvipsnames]{color}
\usepackage[normalem]{ulem}

\begin{document}

\title{Localization of Fe d-states in Ni-Fe-Cu alloys and implications for ultrafast demagnetization}

\author{Ronny Knut} \affiliation{Electromagnetics Division, National Institute of Standards and Technology, Boulder, CO 80305, USA}
\affiliation{Department of Physics and JILA, University of Colorado and NIST, Boulder, CO 80309, USA}

\author{Erna K. Delczeg-Czirjak} 
 \affiliation{Department of Physics and Astronomy, Uppsala University, Box 516, 75120 Uppsala, Sweden}

\author{Justin M. Shaw} 
 \affiliation{Electromagnetics Division, National Institute of Standards and Technology, Boulder, CO 80305, USA}
 
\author{Hans T. Nembach} 
\affiliation{Department of Physics and JILA, University of Colorado and NIST, Boulder, CO 80309, USA}
 \affiliation{Electromagnetics Division, National Institute of Standards and Technology, Boulder, CO 80305, USA}
 
\author{Patrik Grychtol} 
\affiliation{Department of Physics and JILA, University of Colorado and NIST, Boulder, CO 80309, USA}

\author{Dmitriy Zusin} 
\affiliation{Department of Physics and JILA, University of Colorado and NIST, Boulder, CO 80309, USA}

\author{Christian Gentry} 
\affiliation{Department of Physics and JILA, University of Colorado and NIST, Boulder, CO 80309, USA}

\author{Emrah Turgut}
\affiliation{Department of Physics and JILA, University of Colorado and NIST, Boulder, CO 80309, USA}

\author{Henry C. Kapteyn} 
\affiliation{Department of Physics and JILA, University of Colorado and NIST, Boulder, CO 80309, USA}

\author{Margaret M. Murnane} 
\affiliation{Department of Physics and JILA, University of Colorado and NIST, Boulder, CO 80309, USA}

\author{D. A. Arena} 
\affiliation{Department of Physics, University of South Florida, Tampa, FL 33620, USA}
%\affiliation{National Synchrotron Light Source, Brookhaven National Laboratory, Upton, NY 11973, USA}

\author{Olle Eriksson} 
 \affiliation{Department of Physics and Astronomy, Uppsala University, 75120 Uppsala, Sweden}

\author{Olof Karis} 
 \affiliation{Department of Physics and Astronomy, Uppsala University, 75120 Uppsala, Sweden}

\author{T. J. Silva} 
 \affiliation{Electromagnetics Division, National Institute of Standards and Technology, Boulder, CO 80305, USA}

\date{\today}

\begin{abstract}
Ni$_{80}$Fe$_{20}$ (Py) and Py-Cu exhibit intriguing ultrafast demagnetization behavior, where the Ni magnetic moment shows a delayed response relative to the Fe [S. Mathias et al., PNAS {\bf 109}, 4792 (2012)]. To unravel the mechanism responsible for this behavior, we have studied Py-Cu alloys for a wide range of Cu concentrations using X-ray magnetic circular dichroism (XMCD). The magnetic moments of Fe and Ni are found to respond very differently to Cu alloying: Fe becomes a strong ferromagnet in Py, with the magnetic moment  largely unaffected by Cu alloying. In contrast, the Ni magnetic moment decreases continuously with increasing Cu concentration. 
Our results are corroborated by ab-initio calculations of the electronic structure, which we discuss in the framework of virtual bound states (VBSs). 
For high Cu concentrations, Ni exhibits VBSs below the Fermi level, which are likely responsible for an increased orbital/spin magnetic ratio at high Cu concentrations. Fe exhibits VBSs in the minority band, approximately 1 eV above the Fermi level in pure Py, that move closer to the Fermi level upon Cu alloying. A strong interaction between the VBSs and excited electrons above the Fermi level enhances the formation of localized magnons at Fe sites, which explains the different behavior between Fe and Ni during ultrafast demagnetization. 
\end{abstract}

\maketitle

\section{Introduction}

Ni$_{80}$Fe$_{20}$ (Py) is a commonly used material for studying effects involving spin polarized currents, such as spin transfer torque\cite{PhysRevApplied.3.011001}, magnetoresistance\cite{PhysRevB.58.12230, PhysRevApplied.3.011001, PhysRevB.60.477}, spin pumping\cite{PhysRevB.78.014413}, spin hall effect\cite{PhysRevB.89.054401}, inverse spin hall effect\cite{PhysRevLett.111.066602}, and ferromagnetic Josephson junctions\cite{Qader}. It is not only used for its low magnetostriction (no strain induced by magnetization) and the combination of high permeability and low coercivity, but also for its high spin polarization of conduction electrons and large resistivity difference between minority and majority spins\cite{Petrovykh19983459,PhysRevB.60.477}. The origin of this spin dependent resistivity comes mainly from three different effects: magnon scattering, phonon scattering and impurity scattering\cite{PhysRevB.88.184411}. The spin dependent impurity scattering, which has a major contribution even at room temperature, comes from the large impurity potential of Fe in the minority band of Ni\cite{PhysRevB.88.184411,Haidar_2013}.

In a rigid band model (RBM)\cite{Slater1937385}, where the constituents of an alloy are assumed to form a shared band structure, these types of impurity potentials would not exist. 
It has been shown both experimentally\cite{Collins1963633} and theoretically that the RBM is inaccurate for transition metal (TM) alloys, but the behavior of these alloys is anyway described in the spirit of the RBM in many textbooks\cite{Getzlaff20081,Kakehashi,Kittel_book,OHandley_book,Jiles_Book,Berkowitz_Book,Craik_book,Cullity_Book,Tremolet_Book,Vonsovskii_book,Skomski_Book,Bozorth}.
Consequently, the RBM has often been used to explain the Slater-Pauling (SP) curve, 
while the formation of virtual bound states (VBSs) was successfully used for explaining its deviating branches that exhibit antiferromagnetic interactions, such as Cr and V alloyed with Co and Ni\cite{Cable1970176,Dederichs1991241}. The formation of VBSs for dilute alloying concentrations has been found theoretically for a wide range of TM alloys\cite{PhysRevB.49.3352,Jacobs19851941,Vernes2003,PhysRevB.59.419,Choy_aip,PhysRevB.65.075106,Smirnova199914417,PhysRevB.59.419,Paduani_JAP_1999,PhysRevB.50.944,PhysRevB.59.419,Zeller19872123,PhysRevB.49.5157}.
Although, ab-initio calculations have been performed for Py in several publications\cite{PhysRevB.77.054431, PhysRevB.71.134421,Minar_JPhys_2014,PhysRevB.65.075106}, there are basically no discussions about the most prominent feature in its density of states (DOS) and its implications, i.e. the formation of virtual bound states for Fe minority d-states, which form due to the relatively low concentration of Fe (20\%) and the energy mismatch between the minority d-states of Ni and Fe. 

We present x-ray magnetic circular dichroism (XMCD) measurements on a series of (Ni$_{0.8}$Fe$_{0.2}$)$_{1-x}$Cu$_x$ (i.e. Py$_{1-x}$Cu$_{x}$) alloys, where we study the charge transfer and the element specific magnetic moments of Fe and Ni. We show the existence of VBSs in Py$_{1-x}$Cu$_{x}$ by XMCD, and study their behavior using ab-initio calculations. We use these results to shed light on a currently unexplained ultrafast phenomena in Py, where the magnetizations of Fe and Ni show very different responses to excitations by an intense laser pulse. Mathias et al.\cite{Mathias20124792}, found that the Ni magnetization appears to be initially unaffected by the laser pump light and instead follows the demagnetization of the Fe, which responds quickly to the laser light, with an 18 fs delay in Py and 76 fs delay in Py$_{60}$Cu$_{40}$. Considering that pure (not alloyed) Ni and Fe demagnetize on similar time scales, their conclusion might seem surprising. Nevertheless, the results have been reproduced by others\cite{PhysRevB.90.180407, Somnath_JANA} and alternative explanations have not emerged. Importantly, a common misconception is that results by Eschenlohr\cite{EschenlohrThesis} contradict these findings. However, as explicitly explained in Ref.\ \cite{EschenlohrThesis}, a large uncertainty in the determination of the pump-pulse arrival time was handled by shifting the Fe and Ni data sets to eliminate any time-delay between Fe and Ni. 

Fe is found to be a strong ferromagnet in Py, i.e. the Fe majority $d$-states are filled when disregarding $s-p-d$ hybridized states above the Fermi level, and its magnetic moment is largely unaffected by Cu alloying. Ni shows a continuously decreasing magnetic moment with Cu alloying. Ab-initio calculations support these observations and also show the formation of VBSs in the Py DOS. 
Ni has VBS's below the Fermi level for high Cu concentrations and Fe exhibits VBSs in the minority band, $\sim$1 eV above the Fermi level in pure Py, which move closer to the Fermi level with Cu alloying. 
The resulting strong interaction between the VBSs and conduction electrons enhances the exchange interaction between Fe sites. 
Most interestingly, our results support a picture where Elliot-Yafet spin-flip scattering\cite{Koopmans_NatMat_2010, Carva_PRB_2013} is only indirectly responsible for the demagnetization in Py, through an ultrafast generation of magnons. The magnon generation is highly promoted at the Fe sites due to the formation of VBSs and hence the demagnetization of Ni depends on the time it takes for the magnons, which are initially localized at Fe atoms, to distribute throughout the Py lattice. This mechanism for demagnetization is fully consistent with the results and conclusions given by Mathias et al.\cite{Mathias20124792}.

\section{Experimental and theoretical methodology}
Thin films of 20 nm (Fe$_{0.2}$Ni$_{0.8}$)$_{1-x}$Cu$_x$, with $0<x<0.55$ were grown on 3 nm Ta seed layers by magneton sputtering on Si$_3$N$_4$ substrates. The samples were capped by 3 nm Ta to prevent oxidation. Two sample sets, S1 and S2, were grown at two different occasions in the same sputtering system. The results from both sample sets were practically identical, confirming the repeatability of our measurements. Here we present the averaged data of the two samples sets. These samples were grown in the same sputtering system and with the same growth parameters as for the samples studied by Mathias et al.\cite{Mathias20124792} (see their supporting information for sample characterization). 
The XMCD measurements were performed at beamline U4B at the National Synchrotron Light Source at Brookhaven National Laboratory in (Upton, NY), using 70\% circularly polarized light at normal incidence to the samples. The samples were saturated out-of-plane (OOP) using a magnetic field of 1.5 T. The magnetic contrast was obtained by switching the OOP magnetic field, where the corresponding spectra are referred to as M$+$ and M$-$. The measurements were performed in transmission. 
Unlike total electron yield (TEY) measurements, which are more commonly used, the transmission geometry is insensitive to externally applied fields and is not affected by saturation effects which can affect the derived magnetic moments\cite{PhysRevB.59.6421}. Also, transmission measurements provide a much more reliable method for comparing the absorption coefficient between different samples, as they are insensitive to sample dependent electron escape depth and generation of secondary electrons\cite{PhysRevB.59.6421}. A reference sample without the Cu-Py layer was used to correct for the absorption from the Si$_3$N$_4$ and Ta layers.  The provided uncertainties correspond to the standard deviation and are derived by estimating the variance in the difference and sum signal of the M$+$ and M$-$ spectra. Also, an estimated uncertainty in the fitting of the step function was included in the total variance. The accuracy of the film thickness and Cu concentration will affect the analysis of the total absorption cross-section, and was estimated to have a standard deviation of 3\%, as determined from x-ray reflectivity measurements. 

Ab-initio calculations were performed using the spin-polarized relativistic Korringa-Kohn-Rostoker method in combination with the coherent potential approximation (CPA)\cite{0034-4885-74-9-096501, SPRKKR}. To investigate the effect of the exchange-correlation functionals, we used several different implementations of the local density approximation and generalized gradient approximation. We found no significant dependence on the exchange-correlation functional for the quantitative values. Hence, we only present the values obtained from a full potential spin-polarized scalar relativistic calculation with the Perdew-Burke-Ernzerhof functional\cite{PhysRevLett.77.3865} for exchange correlations. 

\section{Experimental results}
\begin{figure}
	\begin{center}
          \includegraphics[width=0.47\textwidth]{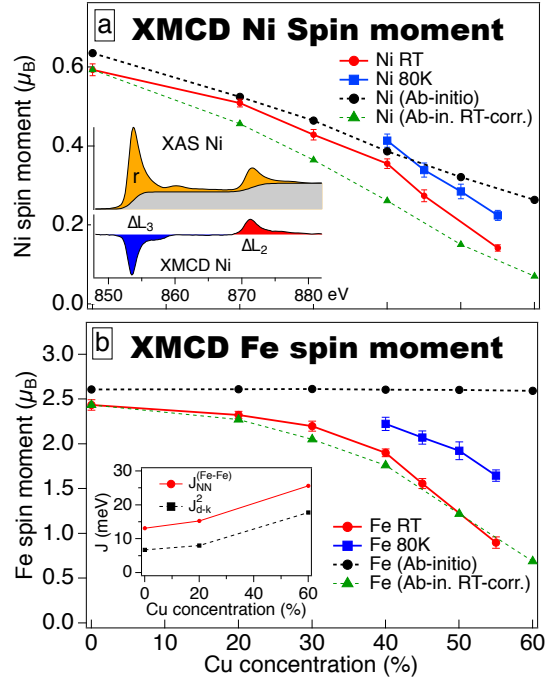}
           \end{center}
      \caption{\label{Spin_FeNi_ratio} Spin magnetic moment of Ni and Fe, respectively, as a function of Cu concentration. The theoretical (black dashed) Ni moment decreases linearly with Cu concentration while the Fe moment remains constant. Inset in a) shows the absorption and XMCD of Ni in Py to illustrate the parameters used in Eq.\ \ref{eq1}. Inset in b) shows the nearest neighbor Fe-Fe exchange interaction ($J_{NN}^{Fe-Fe}$) from ab-initio calculations, and an estimated RKKY-like exchange between magnetic impurities ($J_{d-k}^2$).}
\end{figure}

The spin and orbital magnetic moments are given by the XMCD sum rules\cite{PhysRevLett.75.152} and are proportional to the number of 3d-holes ($N_h$), 
\begin{equation}
\begin{aligned}
m_s=& -\frac{\Delta L_3 - 2 \Delta L_2}{r}\cdot  N_h \\
m_l=& -\frac{2}{3} \frac{\Delta L_3 + \Delta L_2}{r}\cdot  N_h
\label{eq1}
\end{aligned}
\end{equation}
The parameter $r$ is the total integrated intensity of the absorption edge (averaged over both magnetization directions) after subtracting a background (see Fig\ \ref{Spin_FeNi_ratio}a, inset). The position of the steps in the background were taken at the energies corresponding to the peak absorptions of the L$_3$ and L$_2$ edges with a step height ratio of 2:1. $\Delta L_n$ ($n=2,3$) are the integrated absorption differences between M+ and M- spectra at the $L_n$ absorption edges, as illustrated in Fig.\ \ref{Spin_FeNi_ratio}a inset. The magnetic dipole operator can safely be ignored since the sample is thick enough to make any interface effects negligible\cite{PhysRevApplied.2.044014}. The spin moments obtained from XMCD and ab-initio calculations are presented in Fig.\ \ref{Spin_FeNi_ratio}a and b for Ni and Fe, respectively, for different Cu concentrations. The ab-initio spin moments are shown as black dashed/circles, which correspond to a 0 K magnetization. The green dashed/triangles correspond to ab-initio values after correcting for a decreased magnetization at room temperature. The room temperature (RT) correction was obtained from SQUID magnetometry, where we measured the ratio in total magnetization between 300 K and 10 K. Note that this correction is fully valid only when Fe and Ni have equivalent temperature dependence. Since both Fe and Ni show an almost identical XMCD ratio between 300 K and 80 K for 40\% Cu, we can assume this approximation to be reasonable. For deriving the XMCD spin moments, we used the number of holes in Eq.\ \ref {eq1} as a fitting parameter to equate the spin magnetic moment from XMCD and ab-initio at RT for permalloy (Py, 0\% Cu), giving 3.41 holes for Fe and 1.28 holes for Ni in Py.

The spin moments in Py correspond well to values found by neutron scattering techniques \cite{Collins_someTAlloys}, where the magnetic moment for Ni is similar to the pure element ($\sim 0.6 \mu_B$) and the magnetic moment of Fe is enhanced to about $\sim 2.7 \mu_B$ compared to  $\sim 2.2 \mu_B$ in pure bcc Fe.
The theoretical Ni spin moment in Fig.\ \ref{Spin_FeNi_ratio}a shows an almost linear decrease with increasing Cu concentration but remains non-zero even up to 90\% Cu. 
The RT XMCD Ni moment shows values that are somewhat higher than what is obtained from RT-corrected ab-initio values. However, the Ni moment follows a similar trend with Cu concentration. It is possible that the number of Ni $3d$-holes used for deriving the magnetic moment has been overestimated, since a lower number of holes would provide a better correspondence between theory and experiment for Cu alloyed samples. There have been speculations about the mechanism for the decreasing Ni moment in Ni$_{1-x}$Cu$_x$ alloys; however, no clear understanding has yet emerged\cite{PhysRevB.15.1539,Aldred1973218,PhysRevB.25.4937}. Our conclusion is, with the support from the calculated DOS discussed later, that the localization and interaction between Cu and Ni $d$-states drives the reduction of the Ni magnetic moment. 
For the Fe spin moment shown in Fig.\ \ref{Spin_FeNi_ratio}b, theory suggest an almost constant value that persist all the way up to 100\% Cu at $T=0$ K. 
The RT XMCD measurements follows closely the RT-corrected ab-initio values, strongly supporting the 0 K ab-initio result that the local Fe magnetic moment is almost unaffected by Cu alloying. The 80 K magnetization is slightly lower than expected from SQUID measurements. This could possibly be attributed to a higher sample temperature than the temperature reading on the sample holder.

 \begin{figure}
	\begin{center}
          \includegraphics[width=0.53\textwidth]{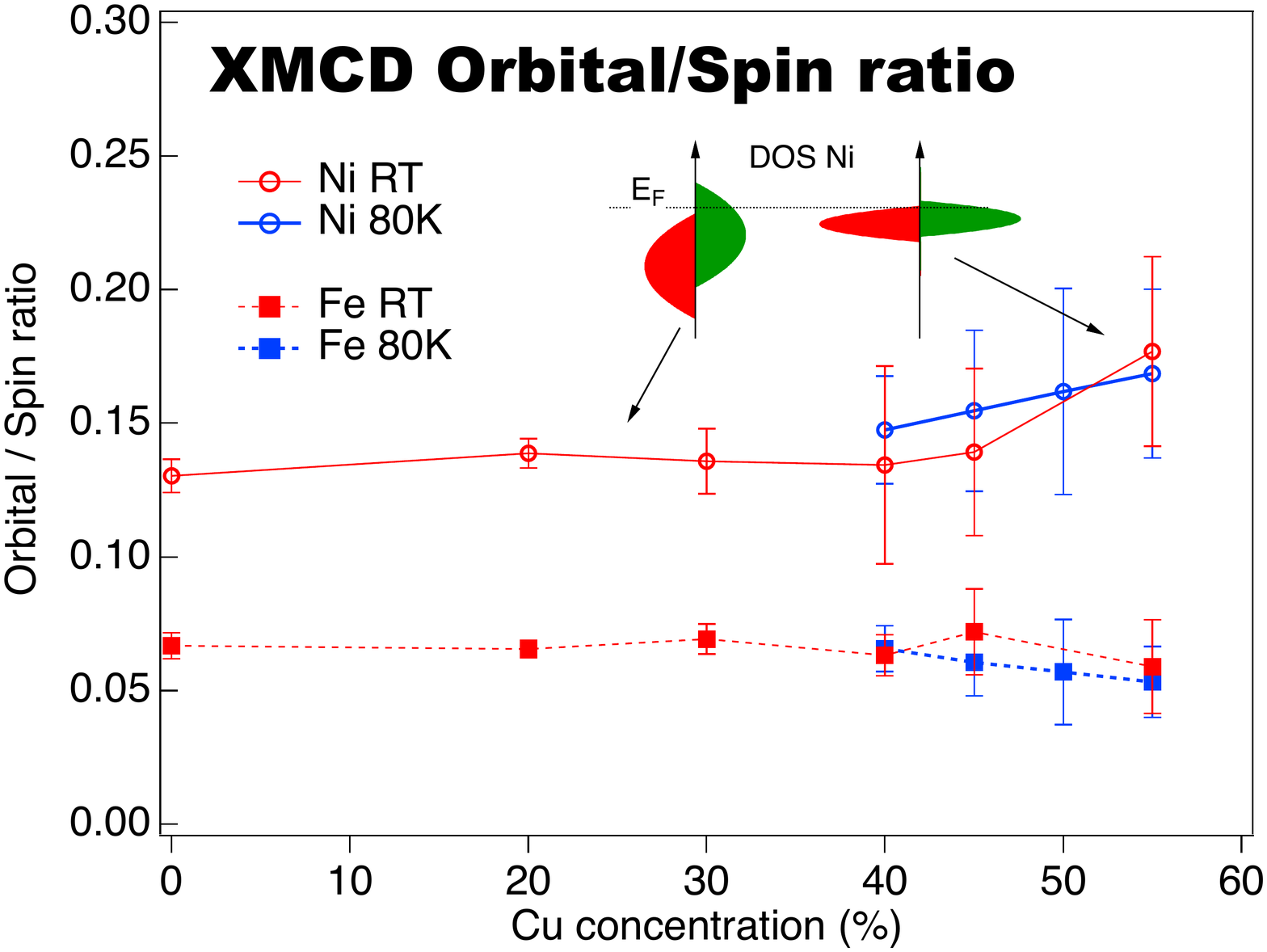}
           \end{center}
      \caption{\label{Spin_Orbit} The ratios between the orbital and spin moments for Fe and Ni in Py$_{1-x}$Cu$_x$ as a function of Cu concentration. The ratio remains constant for low concentrations of Cu alloying. At high Cu concentrations ($>40$\%) the ratio increases for Ni.}
\end{figure}

The orbital moment has a very similar dependence on Cu concentration as the spin magnetic moment.
To identify any Cu concentration dependent trends in the orbital moment, we studied the orbital/spin ratio, shown in Fig.\ \ref{Spin_Orbit}. Unlike the spin and orbital moment by themselves, the $m_l/m_s$ ratio is independent of the number of $3d$-holes, as apparent from Eq.\ \ref{eq1}. The values are slightly higher than what has been measured in other studies for bcc Fe (0.043)\cite{PhysRevLett.75.152}, Ni (0.11)\cite{PhysRevB.53.3409} and Py (Ni: 0.10, Fe: 0.04)\cite{Glaubitz2011}. The ratio for both Fe and Ni remains almost constant through a wide range of Cu concentrations. However, there appears to be an increase in the ratio for Ni at high Cu concentrations. 
This is likely related to the narrowing of occupied Ni $d$-states, see inset in Fig.\ \ref{Spin_Orbit}. This will be discussed later in relation to the calculated density of states.

\begin{figure}
	\begin{center}
          \includegraphics[width=0.47\textwidth]{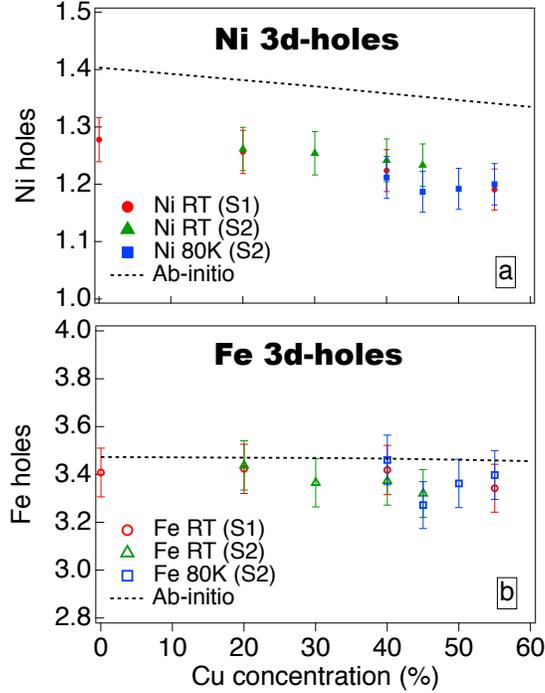}
           \end{center}
      \caption{\label{Spin_Holes} The number of $3d$-holes in a) Ni and b) Fe, obtained from the white-line intensity of the L$_{2,3}$ absorption edges. The number of Fe $3d$ holes appears to be unaffected by Cu alloying while there is a small charge transfer to Ni $3d$-states.}
\end{figure}

The intensity of the L$_{2,3}$ absorption edges corresponds to electron transitions from 2$p_{\frac{1}{2},\frac{3}{2}}$ core levels to empty $3d$ states and is  proportional to the number of $3d$-holes\cite{Stohr1999470}. In Fig.\ \ref{Spin_Holes}a and b, we show the variation in the number of $3d$-holes obtained from the absorption intensity of Ni and Fe, respectively, as a function of Cu concentration. We choose to show both sample sets (S1 and S2) separately since, unlike XMCD, these values are sensitive to variation in both sample thickness and elemental concentrations.
 The numbers of $3d$-holes obtained from our calculations are shown as black dashed lines. The number of holes in Ni, shown in Fig.\ \ref{Spin_Holes}a, obtained from ab-initio is about 10\% higher than what we obtain from the sum rules, but both experiment and theory show a very similar decreasing trend in the number of holes with increased Cu concentration. This indicates either a charge transfer from Cu or a charge reordering between $s-p$ and $d$-states in Ni. In either case, the decrease is very small and can only account for a decreased Ni magnetic moment of 10\% (for 55\% Cu) at most. Similar values have been found experimentally for Ni$_{1-x}$Cu$_x$ alloys\cite{Meitzner1992219} ($\sim 5\%$ decrease of Ni d-holes with 55\% Cu). Theoretically, Stefanou et al.\cite{PhysRevB.35.6911}  has provided a detailed calculation of charge effects for impurities in Ni, which corroborates the weak charge transfer between Cu and Ni. 
The number of holes in Fe, shown in Fig.\ \ref{Spin_Holes}b, gives a much smaller deviation between the XCMD and ab-initio values than Ni. Both theory and experiment indicate an almost constant value of Fe $3d$-holes, independent of the Cu concentration.

The high magnetic moment of Fe, its insensitivity to Cu alloying, and the difference in $3d$-state filling between Fe and Ni, lead us to the following conclusions about the Fe impurities: (1) Since Fe has approximately 2 $\mu_B$ higher magnetic moment than the strong ferromagnet Ni, which corresponds to the difference in d-state filling ($2e^-$, see Fig.\ \ref{Spin_Holes}) between Fe and Ni, then only the filling of the minority band can differ between Fe and Ni in Py. Hence, Fe behaves as a strong ferromagnet, with a filled majority band in Py, regardless of the Cu concentration. (2)  Since Ni and Cu have nearly filled d-bands, the relatively large number of empty minority $d$-states in Fe should not hybridize well with the Ni and Cu $d$-bands. (3) The exchange splitting of the Fe atoms is largely insensitive to the exchange splitting of the Ni-host, as evidenced by the insensitivity of the Fe moment to Cu concentration in contrast to the Ni behavior. From the above statements one can conclude that the minority $d$-states of Fe are decoupled from the host minority $d$-band and could hence be described as virtual bound states above the top of the host $d$-band, as will be described in the following section. 

\section{Virtual bound states (VBS)}

\begin{figure}
	\begin{center}
          \includegraphics[width=0.49\textwidth]{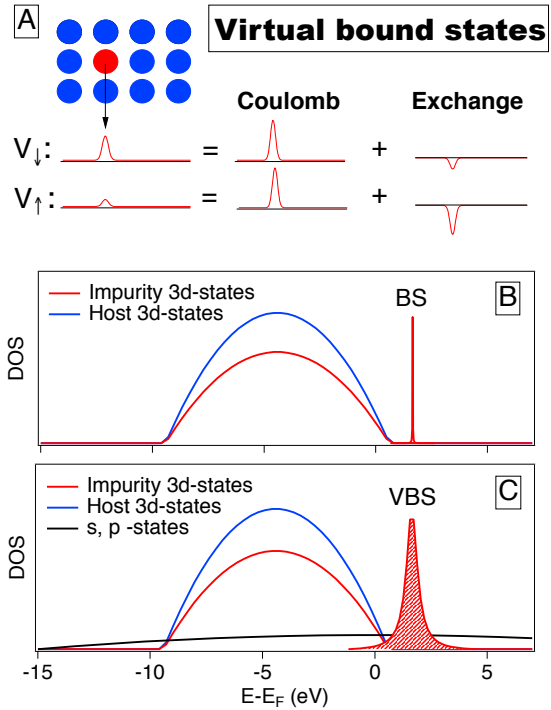}
           \end{center}
      \caption{\label{VBS_Illustration} Illustrations of impurity potentials and virtual bound states. (A) The spin dependent impurity potential is non-zero at impurity sites and can differ between majority ($V_{\uparrow}$)  and minority bands ($V_{\downarrow}$). The impurity potential depends on both the Coulomb and exchange interaction, where the Coulomb contribution is equal for both spin bands while the exchange contribution depends on the occupation number of each spin band. (B) Without hybridization to $s, p$-states, the impurity potential can induce bound states (BS) at the impurity site. (C) The bound states hybridizes with $s, p$-states and form virtual bound states (VBS). This hybridization induces a broadening of the bound states.}
\end{figure}

The valence band of 3d-metals consists of $s$, $p$, and $d$-states. Electrons in $s$ and $p$ states are loosely bound to atoms and can easily form electronic bands, resulting in electrons that are highly mobile. The $d$-states can be considered more localized and less mobile, but their wave functions still extend far enough to hybridize with $d$-states of neighboring atoms and form $d$-bands. However, band formation not only requires a spatial overlap of the wave functions, but also an overlap in momentum space, i.e. the closer electronic states are in both momentum and energy, the stronger they can hybridize. For 3d impurities in a 3d metal host, the free electronic $s$, $p$-states of the host and impurity will easily form a common $s-p$ band. However, the $d$-states will experience a potential change at the impurity atom, where the potential difference between a host and impurity atom is called the impurity potential. As illustrated in Fig.\ \ref{VBS_Illustration}a, the impurity potentials for minority and majority bands, $V_{\downarrow}$ and $V_{\uparrow}$ respectively, depend on both Coulomb and exchange interactions\cite{Campbell1967319}. The Coulomb contribution is the same for both spin bands. However, the exchange interaction depends on the number of occupied states in each band and hence the impurity potentials $V_{\downarrow}$ and $V_{\uparrow}$ can differ. Since the total occupation number is determined by charge neutrality at the impurity site, the impurity potential $V_{\downarrow}$ will decrease if exchange favors a higher $V_{\uparrow}$ potential, and vice versa. 
For a weak impurity potential, hybridization will occur within the $d$-band of the host.
For repulsive impurity potentials i.e. impurities with fewer valence electrons than the host, the density of states for the $3d$-band will increase at higher energies and decrease at lower energies\cite{PhysRev.125.439}.
If the repulsive impurity potential is strong enough, it can even move d-states to energies above the $d$-band, at the impurity site, as illustrated in Fig.\ \ref{VBS_Illustration}b. These states will adopt the character of localized bound states (BS), due to lack of overlap with the host $d$-band\cite{Campbell1967319,Note10d}. 
However, the $s-p$ bands span a very broad energy range. Thus, there is always an energy overlap between $3d$-states and free-electron ($s$, $p$) states. The localized $3d$-states of an impurity will hybridize with free electronic states and can no-longer be considered as bound states, but are referred to as "virtual bound states" (VBS), as illustrated in Fig.\ \ref{VBS_Illustration}c\cite{Anderson196141}. Bound states have a well defined energy but the interaction with free-electronic states provide a Lorentzian energy broadening of the VBSs, indicating the non-infinite lifetime of these states.
 Due to the interaction between the impurity $3d$-states and free-electron states, the VBSs will act as strong scattering centers for conduction electrons if they are close to the Fermi level\cite{Friedel1958287}.

\section{Spin-dependent Density of states}

\begin{figure}
	\begin{center}
          \includegraphics[width=0.49\textwidth]{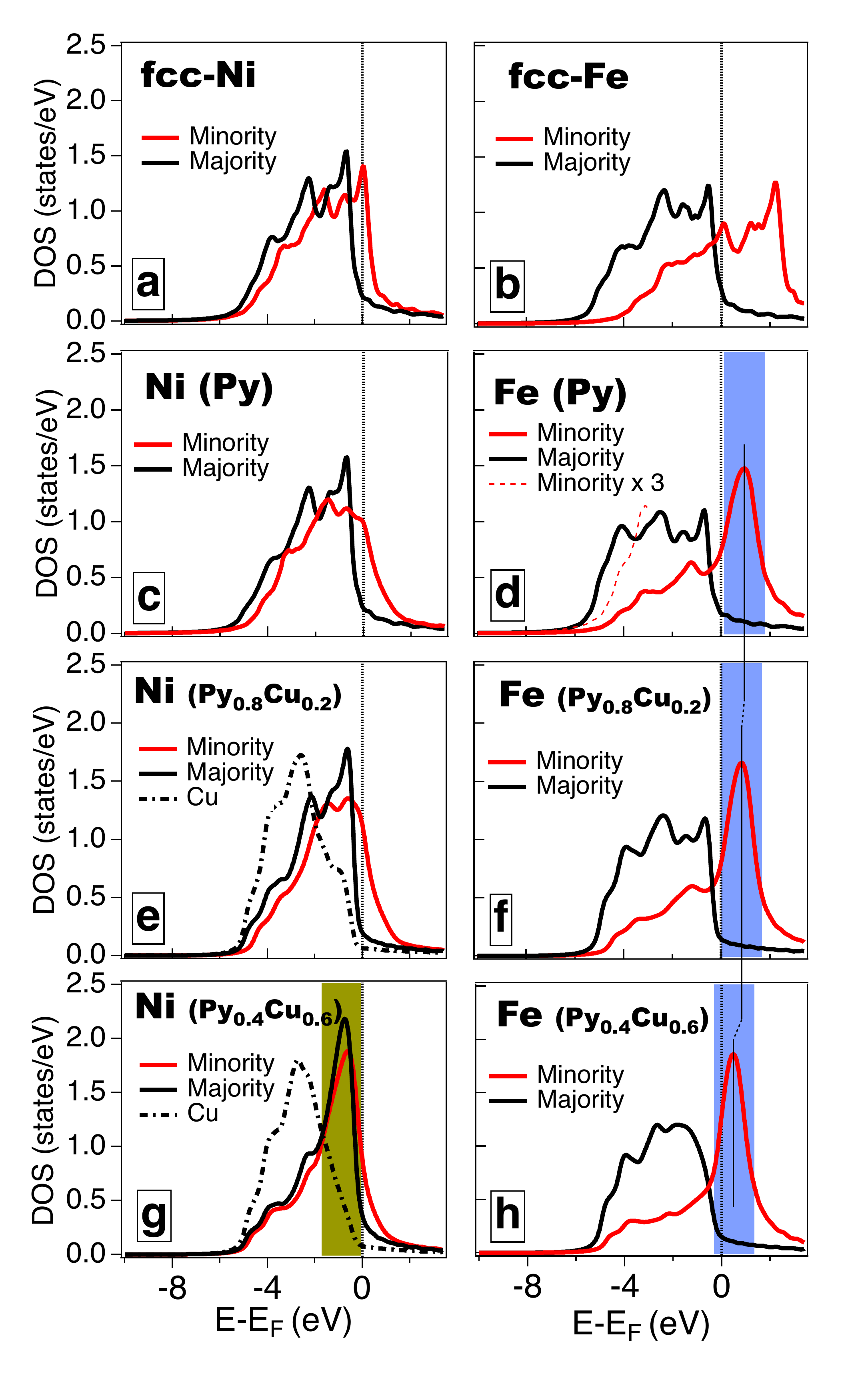}
           \end{center}
      \caption{\label{AbInitio_PyCu_DOS} Spin resolved d-DOS of a) fcc-Ni, b) fcc-Fe, c,e,g) Ni in Py$_{1-x}$Cu$_x$, and d,f,h) Fe in Py$_{1-x}$Cu$_x$ for different Cu concentrations.}
\end{figure}

In Fig.\ \ref{AbInitio_PyCu_DOS} we present the calculated minority and majority $3d$-band DOS for a) fcc-Ni, b) fcc-Fe, c,e,g) Ni in Py$_{1-x}$Cu$_x$, and d,f,h) Fe in Py$_{1-x}$Cu$_x$ for a variety of different Cu concentrations.  Both fcc-Ni and fcc-Fe are found to be strong ferromagnets with almost filled majority bands (Fermi level is indicated by vertical dotted lines). The exchange splitting is 0.7 eV for Ni and 2.5 eV for Fe. Also, the majority and minority bands do not show any significant quantitative differences in structure. In Py, the majority states for both Fe and Ni are similar to the pure elements, but the minority band of Fe is significantly different, see Fig.\ \ref{AbInitio_PyCu_DOS}d. The low energy section (-4 to -6 eV) of the Fe DOS now has a structure and exchange splitting which is similar to Ni as observed by comparing the majority and magnified ($\times 3$) minority band in Fig.\ \ref{AbInitio_PyCu_DOS}d. This is because Fe acts as an impurity in the Ni host band structure with impurity potentials, $V_{\uparrow}$  and $V_{\downarrow}$, for the majority and minority bands, respectively. As discussed in Appendix B, the main part of the impurity potential goes to the minority band for Fe impurities in Ni. 
This is clearly observed in the minority band of Fe, where a large VBS formation is observed, while the disturbance in the majority band is insignificant.

With 20\% Cu in Py, both the width and exchange splitting decreases for majority and minority states in Ni, as shown in Fig.\ \ref{AbInitio_PyCu_DOS}e. 
Since Cu hybridizes with the Ni host and forms a joint $3d$-band (see Cu 3d in Fig.\ \ref{AbInitio_PyCu_DOS}e), it will push Ni $3d$ states to the higher energy side of the $3d$-band while Cu mainly occupies the lower energy section. This type of distribution is a general feature when two states with different energies hybridize\cite{Mohn}, the lower energy states (Cu in this case) will mainly contribute to the lower energy hybridized states, and vice versa. This will also effectively decrease the width of the Ni $3d$-DOS and increase the localization of Ni $3d$ electrons relative to pure Py.  
For higher Cu concentrations (60\%), shown in Fig.\ \ref{AbInitio_PyCu_DOS}g, the localization is so prominent that we observe a well developed VBS below the Fermi level in both majority and minority states of Ni. 
This localization can affect the spin/orbit ratio in various ways and  explain the increase observed for Ni in Fig.\ \ref{Spin_Orbit}. The spin-orbit coupling parameter ($\xi$) increases with decreased orbital size and increased gradient of the atomic potential, which are both possible consequences of VBSs. Also, the crystal field splitting, which can originate from either electrostatic interaction between atoms or from covalent interactions, should decrease due to the diminished hybridization for VBSs\cite{Mohn,PhysRevLett.99.177207}. Any combination of these effects will increase the orbital magnetic moment since in perturbation theory the spin-orbit interaction enters the wave-function effectively as $(\frac{\xi}{\Delta})^n$, where n is the order of the perturbation theory. In this expression
$\Delta$ is the splitting of the energy levels, e.g. due to the crystal field splitting\cite{Stohr_book}. These effects should not be prominent in Fe because all the occupied states are non-localized and only the unoccupied minority states show VBS behavior.
When the Cu concentration increases, the peak in the DOS for the Fe VBS increases, see Fig.\ \ref{AbInitio_PyCu_DOS}g. This is due to an even greater difference in the number of valence electrons between the host and the Fe impurity that results in an increased $V_{\downarrow}$. The black solid vertical line that cuts through Figs.\ \ref{AbInitio_PyCu_DOS}d, f, and h, shows the energy shift of the VBS peak position. The peak position should move to higher energies with increased $V_{\downarrow}$. However, this effect is compensated by a shift of the host minority $d$-band to lower energies due to high Cu content and a reduced Ni exchange splitting. The end result is that the Fe minority VBS peak shifts 0.4 eV lower in energies as the Cu concentration increases from 0\% to 60\%.

The exchange splitting for Ni $3d$-states vanishes at high Cu concentrations while it remains large for Fe. This effect was explained by Anderson \cite{Anderson196141}, who showed that the fractional filling of the $d$-states together with the Coulomb and exchange interaction determines if the ground state is magnetic. In Appendix A, we use the Anderson impurity model to demonstrate that Ni and Fe should indeed be non-magnetic and magnetic, respectively, in a Cu host.

\section{Free electron-VBS interaction}
An important consequence of the VBS formation is the strong interaction with free electron states. In the inset of Fig.\ \ref{Spin_FeNi_ratio}b, we show the dependence of the exchange interaction upon Cu concentration between two Fe atom when they are nearest neighbors, $J_{NN}^{Fe-Fe}$,  as obtained from our ab-initio calculations. Unlike all other exchange interactions between Fe-Ni and Ni-Ni that decreases with Cu concentration, $J_{NN}^{Fe-Fe}$ increases strongly. This is a direct result of VBS formation in the Fe minority band and their ability to be strong scattering centers for electrons in the minority $s-p$ band. This scattering results in a strong exchange interaction between the localized Fe $3d$-magnetic moments and the free electrons. As we have shown in the previous section, Cu alloying increases the density of VBSs and moves them closer to the Fermi level. Therefore, the scattering of conduction electrons at the Fermi level and their exchange interaction with Fe $3d$ states also increases with Cu concentration. This RKKY-like exchange will couple Fe atoms with each other as evidenced in the increasing  nearest neighbor exchange interaction ($J_{NN}^{Fe-Fe}$), which is shown in the inset of Fig.\ \ref{Spin_FeNi_ratio}b. In the inset, we also show how the RKKY-like pair exchange interaction is expected to depend on the Cu concentration in the Anderson impurity model ($J_{d-k}^2\sim \frac{1}{(\epsilon_k-\epsilon_d)^2}$), where $\epsilon_d$ is the peak energy of the VBS and $\epsilon_k$ is the energy of the free electrons\cite{PhysRev.149.491, Schrieffer19671143, PhysRev.106.893}. This model is described further in Appendix B. Note that this model assumes highly dilute localized moments in a non-magnetic host. As such, we only show it here to illustrate that the general trend of $J_{NN}^{Fe-Fe}$ is almost identical to what is expected from an RKKY-like exchange. 
Experimentally, it has been shown that FeCu alloys can indeed exhibit spin-glass behavior\cite{PhysRevB.33.3247}, which is indicative of such an RKKY-like exchange interactions\cite{Mohn}. 

\section{Discussion}
 
We have discussed impurity potentials in the framework of virtual bound states, and shown that the element specific magnetization of Py obtained from XMCD is consistent with the ab-initio prediction of VBS formation at Fe sites in the minority band. We expect that these states should be the main source of impurity scattering in Py, and can be controlled by adjusting the Cu concentration, as was shown by our ab-initio calculations. Theory predicts that the density of VBSs at the Fermi level increases with Cu alloying. This should enhance the conductivity difference between the majority band and minority band, even though the overall reduction in magnetization would certainly counteract this effect. 

We have shown that the exchange interaction between Fe impurities strongly increases with Cu alloying. This comes from an increased exchange coupling between localized Fe magnetic moments and $s-p$ states at the Fermi level. The VBSs in Py are approximately 1 eV above the Fermi level. This suggests that electrons excited with an 800 nm wavelength ($\sim$1.5 eV) laser pulse should interact strongly with VBSs in Py, resulting in a large exchange interaction between excited electrons and localized Fe moments. 
\begin{figure}
	\begin{center}
          \includegraphics[width=0.49\textwidth]{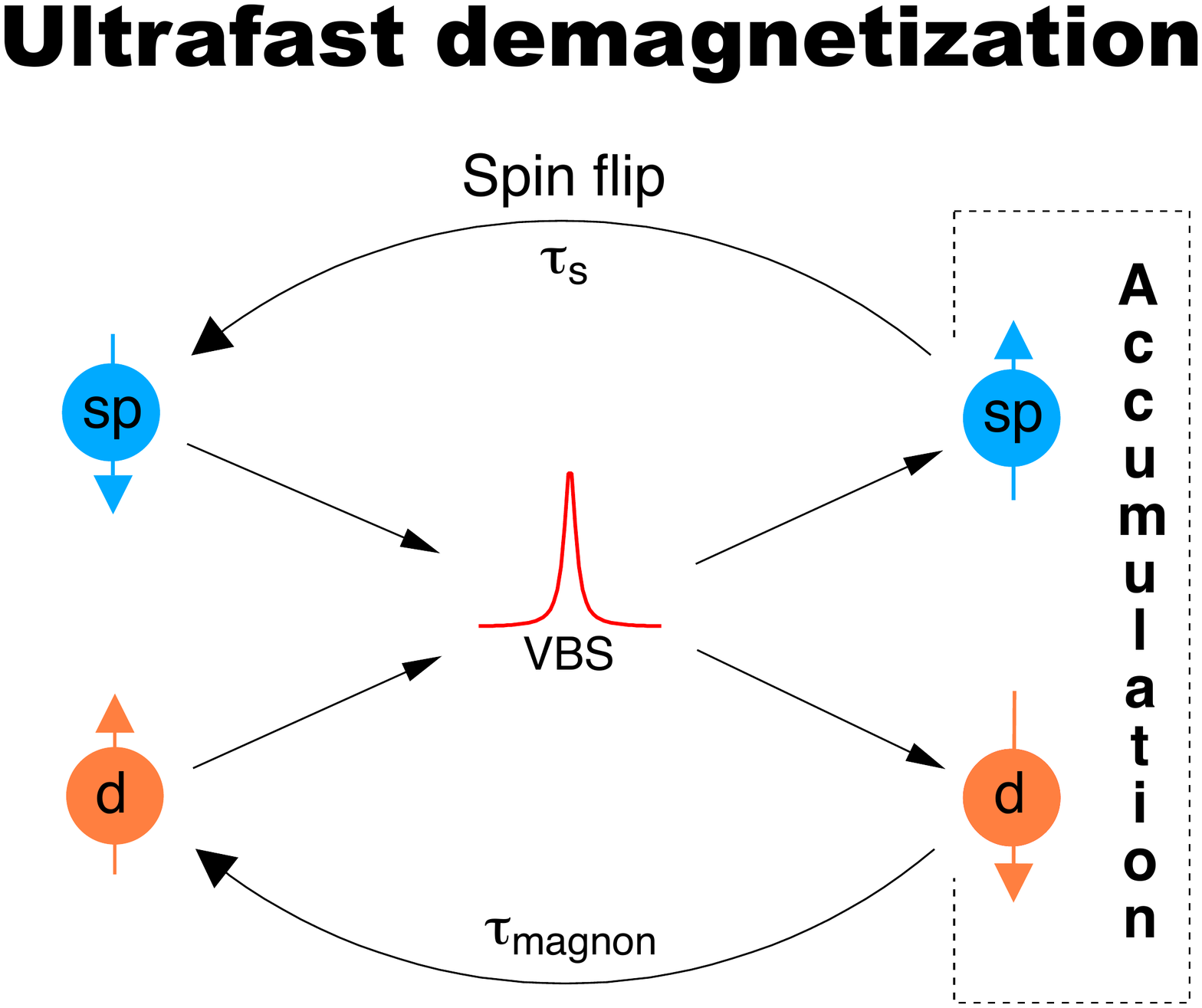}
           \end{center}
      \caption{\label{SpinFlipMagnon} Illustration of magnon generation and spin flip processes during demagnetization of Py. Free-electrons in the minority band will quickly scatter on VBSs and generate magnons. Electrons in the majority band can go to the minority band through a spin-flip process, which is responsible for the angular momentum transfer to the lattice. The process can repeat until the electrons reach their ground state.}
\end{figure}
As discussed by Haag et al.\cite{PhysRevB.90.014417, Haag2013, PhysRevB.88.214404}, the electron-magnon scattering rate depends on the exchange interaction between free electrons and the localized magnetic moments. {\it Hence we surmise that, the magnon emission rate should strongly increase at Fe sites for electrons excited $\sim1$ eV above the Fermi level.}
Haag et al.\cite{PhysRevB.90.014417}, also showed that the magnon generation is much faster in pure elemental Fe, as compared to Ni, and the magnon emission rate is higher than the Elliot-Yafet spin flip rate for both elements. They suggest that, even though magnon emission does not change the magnetization due to the concurrent spin-flip process, it promotes s-p electrons from the minority to the majority band, which eventually decreases the magnetization via electron-phonon scattering. Similar magnon generation has also been compared to pure spin-flip excitations by Schellekens and Koopmans\cite{PhysRevLett.110.217204}, where they found an enhanced demagnetization rate in the former case. This suggests that a qualitatively accurate description of ultrafast demagnetization, must include the effect of electron-magnon scattering.

Enhanced magnon generation at Fe sites should strongly influence the ultrafast demagnetization processes for Py and Py-Cu alloys. As mentioned in the introduction, Mathias et al.\cite{Mathias20124792} observed a significant delay between Fe and Ni demagnetization in Py$_{1-x}$Cu$_{x}$. The Cu alloying decreases the average exchange interaction between magnetic elements in Py and increases the delay of the Ni demagnetization. 
In Fig.\ \ref{SpinFlipMagnon} we illustrate the proposed demagnetization process in Py-Cu alloys. In the minority band, free electrons will rapidly scatter on VBSs, and generate magnons (represented by a flipped d-electron), while populating the majority band. When this magnon generation is faster than the spin-flip time ($\tau_s$), there will be an accumulation of majority spins. This imbalance between majority and minority spins will enhance the demagnetization rate by Elliot-Yafet spin-flip scattering, which is responsible for the transfer of angular momentum from the spins to the lattice. Since this process will effectively increase the magnetization of the free-electrons, the resulting demagnetization is purely a result of magnon generation. Therefore it is critical that the magnon decay time ($\tau_{magnon}$) is long enough to accumulate magnons. Commonly, $\tau_{magnon}=1/ (\alpha \cdot 4\pi f)$, where $\alpha \approx 0.01$ is the damping parameter and $f\approx 1$ THz is the frequency region for short wave length magnons\cite{PhysRevB.11.2668}. This gives $\tau_{magnon} \approx 8$ ps, which is more than an order of magnitude longer than demagnetization times in Py-Cu. Note that the free electrons which transfer the angular momentum to the lattice, can repeat this process until they have lost their energy and reaches the Fermi level, further enhancing the demagnetization rate of the sample.

Since the magnons will initially be localized at Fe sites, ultrafast demagnetization should initially proceed in the Fe sublattice. It is not within the scope of this paper to provide a model of the magnon dynamics. However, we can make the following statement: 
 The group velocity of long wave length magnons, $v_g\approx 2Dk/\hbar$, is proportional to the average exchange interaction, i.e. $D \propto J_{Avg}$, where $J_{Avg}$ is the average pair exchange. Thus the time it takes for the magnons, with typical group velocities on the order of $10^5$ m/s for exchange magnons, to distribute throughout the lattice is inversely proportional to the exchange interaction that decreases with increased Cu concentration, consistent with the observations of 20-80 fs delay between Fe and Ni demagnetizations as found by Mathias et al.\cite{Mathias20124792}.
If the magnon generation rate is higher than the Elliot-Yafet spin-flip rate in pure Fe and Ni, as found by Haag et al.\cite{PhysRevB.90.014417}, then the impact of VBSs for the non-elemental resolved demagnetization rate should be negligible and the effect of VBSs should only manifest as a demagnetization delay between the different elements. 

\section{Conclusions}
 
We have have shown that the decreased magnetization of Py, when alloyed with Cu, is the result of Ni $d$-state localization, in addition to the hybridization of Ni and Cu $d$-bands. 
Only a small effect from $d$-band filling ($\sim$6\% decrease of Ni $3d$-holes with 55\% Cu alloying), as suggested by the RBM, is observed.
Fe in Py and Py-Cu forms VBSs in the minority band above the Fermi level. These states ensure that the Fe majority band remains filled for all Cu concentrations, resulting in a high Fe moment of 2.6 $\mu_B$. 
The intensity of the VBS increases with increasing Cu concentration since the average number of host valence electrons increases. 
The increased VBS intensity at the Fermi level increases the scattering of minority conduction electrons.
 Also, this scattering of free electrons on VBS states causes a strong exchange coupling between Fe sites, similar to an RKKY interaction. This is observed as an increase in the Fe-Fe exchange interaction obtained from ab-initio calculations. 
In the discussion section we propose a mechanism for ultrafast demagnetization in Py and Py-Cu, which explains the different responses found for Fe and Ni. The VBSs should have a significant impact on the probability of generating magnons, which are initially localized at Fe sites. The time-delay between the Fe and Ni demagnetization is proportional to the time it takes for these magnons to distribute throughout the Py lattice. The magnon generation also creates an imbalance in occupation of the spin-dependent $s-p$ bands, which enhances the demagnetization rate.

\section{Acknowledgements}
We thank Yves Idzerda for useful discussions.
This work was partially supported by the U.S Department of Energy Office of Basic Energy Sciences X-Ray Scattering Program (Grants \#DE-SC0002002 and \#DE-FG02-09ER46652).
O.E. acknowledges support from the KAW foundation and VR. R.K. acknowledges the Swedish Research Council (VR) for their financial support. E. K. D.-Cz. acknowledges National Supercomputer Centre at Link{\"o}ping University Sweden for computational resources. P.G. acknowledges support from the Deutsche Forschungsgemeinschaft (no. GR 4234/1-1). Use of the National Synchrotron Light Source, Brookhaven National Laboratory, was supported by the U.S. Department of Energy, Office of Science, Office of Basic Energy Sciences, under Contract No. DE-AC02-98CH10886.

\section{Appendix A: Anderson impurity model and exchange interaction}

Since the VBS formation of Fe is significantly above the $d$-band of Ni, it is justifiable to discuss VBSs using Andersons model, even though the model is only strictly applicable for magnetic atoms in a non-magnetic host\cite{Anderson196141}.
The Anderson impurity model describes the mixing between bound states and free electrons, rather than the actual formation of the bound states. The half width at half maximum ($\Delta$) of the VBS, in this model, is a direct consequence of the mixing between $d-$ and free electron states, $\Delta=\pi\left<V_{kd}^2\right>_{av}\rho(\epsilon)$, where $V_{kd}$ is the mixing parameter and $\rho(\epsilon)$ is the free electron density at the VBS energy position.\cite{Anderson196141}  Using $\Delta=0.7\ eV$ and $\rho(\epsilon)=0.046\ states/eV/atom$, as found from our ab-initio calculations for Py$_{0.8}$Cu$_{0.2}$, we obtain a mixing parameter of $V_{kd}=2.2\ eV$. This is in the region (2-3 eV) that was predicted by Anderson for 3d-metals. Also, the width remains almost constant with Cu alloying, which is consistent with a nearly constant $s,p$ DOS in the VBS energy region, found by our ab-initio calculations.

According to Anderson\cite{Anderson196141}, for magnetic impurities in a non-magnetic metal, the condition for a magnetic state is $\frac{J+2U}{\Delta}\geq \frac{2\pi}{\sin^2(\pi\cdot n_c)}$, where U and J are the effective Coulomb and exchange interactions, respectively. We have used $U=4\ eV$ and $J=0.75\ eV$ which are appropriate for Ni in Anderson impurity model\cite{PhysRevB.83.121101,PhysRevB.85.085114}. The fractional filling of the $d$-band ($n_c$) is directly given by the number of holes in the non-magnetic state ($n_c=0.66$ for Fe and 0.87 for Ni). We find that Fe clearly satisfies the condition for band splitting ($12.5\geq8$) while Ni is far from satisfying it ($12.5\ngeq40$). Note that $\Delta$ is likely broadened by the CPA treatment of the DOS as well as the crystal field splittings, however, even a 3 times smaller $\Delta$ will keep the Fe spin split while Ni becomes non-magnetic.

In the context of Anderson theory it is possible to estimate the exchange interaction between magnetic impurities in a metallic host. In this model the exchange interaction between impurities is mediated by a RKKY-like interaction.
Since VBSs are strong scattering centers for conduction electrons they can induce a spin polarization of the latter. 
This polarization will mediate the effective pair exchange coupling between the localized moments.
 The exchange between a localized moment and conduction electrons can be approximated by $J_{d-k}=\frac{\mid v_{kd} \mid^2}{2S} \frac{U}{\epsilon^d(\epsilon^d-U)}$\cite{Schrieffer19671143}
, where $\epsilon^d$ is the energy of the VBS peak relative to the Fermi level, S is the spin of the Fe impurity. 
However, if one assumes that the scattering electrons ($\epsilon_k$) are above the Fermi level and also that there are only VBSs in one of the spin bands, then $J_{d-k}=\frac{\mid v_{kd} \mid^2}{2S} \frac{1}{\epsilon_k-\epsilon_d}$.
The interaction between Fe moments is then proportional to $(J_{d-k})^2$\cite{PhysRev.106.893}.

\section{Appendix B: Charge displacement}

\begin{figure}
	\begin{center}
          \includegraphics[width=0.52\textwidth]{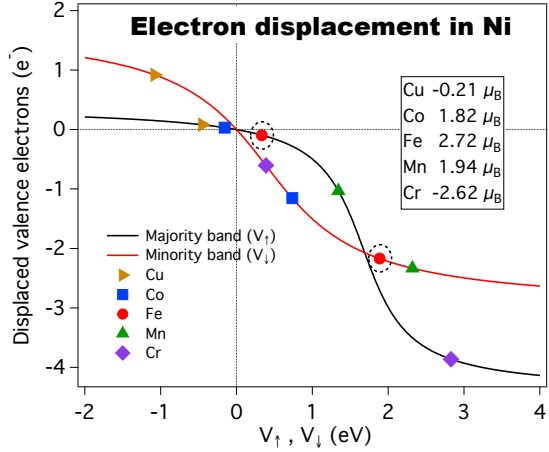}
           \end{center}
      \caption{\label{Charge_Disp} The electron displacement for both minority and majority bands as functions of V$_{\downarrow}$ and V$_{\uparrow}$, respectively. The electron displacement for different impurities (Cu, Co, Fe, Mn and Cr) in a Ni host, is indicated. The corresponding magnetic moments of the impurities are also listed.}
\end{figure}

There are essentially two types of VBSs. The first corresponds to localized states of impurities in a free electron metal, e.g. Fe impurity in Al host, which is the case described by Anderson\cite{Anderson196141}. The second case corresponds to changing the energy and behavior of localized states that already exists in a host (the $3d$ band) by adding an impurity, e.g. Fe impurity in Ni host, which can be described by the impurity potentials $V_{\downarrow}$ and $V_{\uparrow}$ in the minority and majority bands, respectively. In general, the impurity potential is determined by the Coulomb and intra-exchange energies of the impurity atom\cite{Campbell1967319}. 
The distribution of  this impurity potential between majority and minority bands is strongly dependent on the difference in the number of valence electrons between the host and impurity. In turn, this difference will determine if the  impurity will be ferromagnetically or anti-ferromagnetically aligned to the host.
By using the formalism of Campbell and Gomes\cite{Campbell1967319}, and Clogston\cite{PhysRev.125.439}, it is possible to use the Ni host DOS and the valence electron difference between the host and impurity to self-consistently solve for the impurity potential distribution. In Fig.\ \ref{Charge_Disp}, we present such calculations for various impurities in Ni.
We show these calculations solely for illustrative purposes.
 The calculations are identical to what was done by Campbell and Gomes\cite{Campbell1967319}, except that a more realistic DOS obtained via ab-initio calculations was used. The charge displacement on the left axis corresponds to how much charge has been displaced from below the Fermi level to above it due to the impurity potentials $V_{\downarrow}$ and  $V_{\uparrow}$ in the minority and majority band, respectively. For the majority band, there is a threshold before the impurity potential $V_{\uparrow}$ can displace any electrons. For sufficiently large impurity potentials, the majority band can displace more electrons than the minority band. This is why Fe and Co, which have a small difference in the number of valence electrons relative to the Ni host, will only push minority states above the Fermi level. For larger electron displacements, as for Cr, the impurity potential mainly goes to the majority band, which makes Cr antiferrromagnetically aligned with the Ni host. (Note that quantitative DFT calculations\cite{Zeller19872123} predict a substantial impurity potential in the minority band as well for Cr).

\bibliography{Refs_CuPy}

\begin{thebibliography}{80}%
\makeatletter
\providecommand \@ifxundefined [1]{%
 \@ifx{#1\undefined}
}%
\providecommand \@ifnum [1]{%
 \ifnum #1\expandafter \@firstoftwo
 \else \expandafter \@secondoftwo
 \fi
}%
\providecommand \@ifx [1]{%
 \ifx #1\expandafter \@firstoftwo
 \else \expandafter \@secondoftwo
 \fi
}%
\providecommand \natexlab [1]{#1}%
\providecommand \enquote  [1]{``#1''}%
\providecommand \bibnamefont  [1]{#1}%
\providecommand \bibfnamefont [1]{#1}%
\providecommand \citenamefont [1]{#1}%
\providecommand \href@noop [0]{\@secondoftwo}%
\providecommand \href [0]{\begingroup \@sanitize@url \@href}%
\providecommand \@href[1]{\@@startlink{#1}\@@href}%
\providecommand \@@href[1]{\endgroup#1\@@endlink}%
\providecommand \@sanitize@url [0]{\catcode `\\12\catcode `\$12\catcode
  `\&12\catcode `\#12\catcode `\^12\catcode `\_12\catcode `\%12\relax}%
\providecommand \@@startlink[1]{}%
\providecommand \@@endlink[0]{}%
\providecommand \url  [0]{\begingroup\@sanitize@url \@url }%
\providecommand \@url [1]{\endgroup\@href {#1}{\urlprefix }}%
\providecommand \urlprefix  [0]{URL }%
\providecommand \Eprint [0]{\href }%
\providecommand \doibase [0]{http://dx.doi.org/}%
\providecommand \selectlanguage [0]{\@gobble}%
\providecommand \bibinfo  [0]{\@secondoftwo}%
\providecommand \bibfield  [0]{\@secondoftwo}%
\providecommand \translation [1]{[#1]}%
\providecommand \BibitemOpen [0]{}%
\providecommand \bibitemStop [0]{}%
\providecommand \bibitemNoStop [0]{.\EOS\space}%
\providecommand \EOS [0]{\spacefactor3000\relax}%
\providecommand \BibitemShut  [1]{\csname bibitem#1\endcsname}%
\let\auto@bib@innerbib\@empty
%</preamble>
\bibitem [{\citenamefont {Baek}\ \emph {et~al.}(2015)\citenamefont {Baek},
  \citenamefont {Rippard}, \citenamefont {Pufall}, \citenamefont {Benz},
  \citenamefont {Russek}, \citenamefont {Rogalla},\ and\ \citenamefont
  {Dresselhaus}}]{PhysRevApplied.3.011001}%
  \BibitemOpen
  \bibfield  {author} {\bibinfo {author} {\bibfnamefont {B.}~\bibnamefont
  {Baek}}, \bibinfo {author} {\bibfnamefont {W.~H.}\ \bibnamefont {Rippard}},
  \bibinfo {author} {\bibfnamefont {M.~R.}\ \bibnamefont {Pufall}}, \bibinfo
  {author} {\bibfnamefont {S.~P.}\ \bibnamefont {Benz}}, \bibinfo {author}
  {\bibfnamefont {S.~E.}\ \bibnamefont {Russek}}, \bibinfo {author}
  {\bibfnamefont {H.}~\bibnamefont {Rogalla}}, \ and\ \bibinfo {author}
  {\bibfnamefont {P.~D.}\ \bibnamefont {Dresselhaus}},\ }\href {\doibase
  10.1103/PhysRevApplied.3.011001} {\bibfield  {journal} {\bibinfo  {journal}
  {Phys. Rev. Applied}\ }\textbf {\bibinfo {volume} {3}},\ \bibinfo {pages}
  {011001} (\bibinfo {year} {2015})}\BibitemShut {NoStop}%
\bibitem [{\citenamefont {Holody}\ \emph {et~al.}(1998)\citenamefont {Holody},
  \citenamefont {Chiang}, \citenamefont {Loloee}, \citenamefont {Bass},
  \citenamefont {Pratt},\ and\ \citenamefont {Schroeder}}]{PhysRevB.58.12230}%
  \BibitemOpen
  \bibfield  {author} {\bibinfo {author} {\bibfnamefont {P.}~\bibnamefont
  {Holody}}, \bibinfo {author} {\bibfnamefont {W.~C.}\ \bibnamefont {Chiang}},
  \bibinfo {author} {\bibfnamefont {R.}~\bibnamefont {Loloee}}, \bibinfo
  {author} {\bibfnamefont {J.}~\bibnamefont {Bass}}, \bibinfo {author}
  {\bibfnamefont {W.~P.}\ \bibnamefont {Pratt}}, \ and\ \bibinfo {author}
  {\bibfnamefont {P.~A.}\ \bibnamefont {Schroeder}},\ }\href {\doibase
  10.1103/PhysRevB.58.12230} {\bibfield  {journal} {\bibinfo  {journal} {Phys.
  Rev. B}\ }\textbf {\bibinfo {volume} {58}},\ \bibinfo {pages} {12230}
  (\bibinfo {year} {1998})}\BibitemShut {NoStop}%
\bibitem [{\citenamefont {Dubois}\ \emph {et~al.}(1999)\citenamefont {Dubois},
  \citenamefont {Piraux}, \citenamefont {George}, \citenamefont {Ounadjela},
  \citenamefont {Duvail},\ and\ \citenamefont {Fert}}]{PhysRevB.60.477}%
  \BibitemOpen
  \bibfield  {author} {\bibinfo {author} {\bibfnamefont {S.}~\bibnamefont
  {Dubois}}, \bibinfo {author} {\bibfnamefont {L.}~\bibnamefont {Piraux}},
  \bibinfo {author} {\bibfnamefont {J.~M.}\ \bibnamefont {George}}, \bibinfo
  {author} {\bibfnamefont {K.}~\bibnamefont {Ounadjela}}, \bibinfo {author}
  {\bibfnamefont {J.~L.}\ \bibnamefont {Duvail}}, \ and\ \bibinfo {author}
  {\bibfnamefont {A.}~\bibnamefont {Fert}},\ }\href {\doibase
  10.1103/PhysRevB.60.477} {\bibfield  {journal} {\bibinfo  {journal} {Phys.
  Rev. B}\ }\textbf {\bibinfo {volume} {60}},\ \bibinfo {pages} {477} (\bibinfo
  {year} {1999})}\BibitemShut {NoStop}%
\bibitem [{\citenamefont {Ando}\ \emph {et~al.}(2008)\citenamefont {Ando},
  \citenamefont {Kajiwara}, \citenamefont {Takahashi}, \citenamefont {Maekawa},
  \citenamefont {Takemoto}, \citenamefont {Takatsu},\ and\ \citenamefont
  {Saitoh}}]{PhysRevB.78.014413}%
  \BibitemOpen
  \bibfield  {author} {\bibinfo {author} {\bibfnamefont {K.}~\bibnamefont
  {Ando}}, \bibinfo {author} {\bibfnamefont {Y.}~\bibnamefont {Kajiwara}},
  \bibinfo {author} {\bibfnamefont {S.}~\bibnamefont {Takahashi}}, \bibinfo
  {author} {\bibfnamefont {S.}~\bibnamefont {Maekawa}}, \bibinfo {author}
  {\bibfnamefont {K.}~\bibnamefont {Takemoto}}, \bibinfo {author}
  {\bibfnamefont {M.}~\bibnamefont {Takatsu}}, \ and\ \bibinfo {author}
  {\bibfnamefont {E.}~\bibnamefont {Saitoh}},\ }\href {\doibase
  10.1103/PhysRevB.78.014413} {\bibfield  {journal} {\bibinfo  {journal} {Phys.
  Rev. B}\ }\textbf {\bibinfo {volume} {78}},\ \bibinfo {pages} {014413}
  (\bibinfo {year} {2008})}\BibitemShut {NoStop}%
\bibitem [{\citenamefont {Niimi}\ \emph {et~al.}(2014)\citenamefont {Niimi},
  \citenamefont {Suzuki}, \citenamefont {Kawanishi}, \citenamefont {Omori},
  \citenamefont {Valet}, \citenamefont {Fert},\ and\ \citenamefont
  {Otani}}]{PhysRevB.89.054401}%
  \BibitemOpen
  \bibfield  {author} {\bibinfo {author} {\bibfnamefont {Y.}~\bibnamefont
  {Niimi}}, \bibinfo {author} {\bibfnamefont {H.}~\bibnamefont {Suzuki}},
  \bibinfo {author} {\bibfnamefont {Y.}~\bibnamefont {Kawanishi}}, \bibinfo
  {author} {\bibfnamefont {Y.}~\bibnamefont {Omori}}, \bibinfo {author}
  {\bibfnamefont {T.}~\bibnamefont {Valet}}, \bibinfo {author} {\bibfnamefont
  {A.}~\bibnamefont {Fert}}, \ and\ \bibinfo {author} {\bibfnamefont
  {Y.}~\bibnamefont {Otani}},\ }\href {\doibase 10.1103/PhysRevB.89.054401}
  {\bibfield  {journal} {\bibinfo  {journal} {Phys. Rev. B}\ }\textbf {\bibinfo
  {volume} {89}},\ \bibinfo {pages} {054401} (\bibinfo {year}
  {2014})}\BibitemShut {NoStop}%
\bibitem [{\citenamefont {Miao}\ \emph {et~al.}(2013)\citenamefont {Miao},
  \citenamefont {Huang}, \citenamefont {Qu},\ and\ \citenamefont
  {Chien}}]{PhysRevLett.111.066602}%
  \BibitemOpen
  \bibfield  {author} {\bibinfo {author} {\bibfnamefont {B.~F.}\ \bibnamefont
  {Miao}}, \bibinfo {author} {\bibfnamefont {S.~Y.}\ \bibnamefont {Huang}},
  \bibinfo {author} {\bibfnamefont {D.}~\bibnamefont {Qu}}, \ and\ \bibinfo
  {author} {\bibfnamefont {C.~L.}\ \bibnamefont {Chien}},\ }\href {\doibase
  10.1103/PhysRevLett.111.066602} {\bibfield  {journal} {\bibinfo  {journal}
  {Phys. Rev. Lett.}\ }\textbf {\bibinfo {volume} {111}},\ \bibinfo {pages}
  {066602} (\bibinfo {year} {2013})}\BibitemShut {NoStop}%
\bibitem [{\citenamefont {Abd El~Qader}\ \emph {et~al.}(2014)\citenamefont {Abd
  El~Qader}, \citenamefont {Singh}, \citenamefont {Galvin}, \citenamefont {Yu},
  \citenamefont {Rowell},\ and\ \citenamefont {Newman}}]{Qader}%
  \BibitemOpen
  \bibfield  {author} {\bibinfo {author} {\bibfnamefont {M.}~\bibnamefont {Abd
  El~Qader}}, \bibinfo {author} {\bibfnamefont {R.~K.}\ \bibnamefont {Singh}},
  \bibinfo {author} {\bibfnamefont {S.~N.}\ \bibnamefont {Galvin}}, \bibinfo
  {author} {\bibfnamefont {L.}~\bibnamefont {Yu}}, \bibinfo {author}
  {\bibfnamefont {J.~M.}\ \bibnamefont {Rowell}}, \ and\ \bibinfo {author}
  {\bibfnamefont {N.}~\bibnamefont {Newman}},\ }\href {\doibase
  http://dx.doi.org/10.1063/1.4862195} {\bibfield  {journal} {\bibinfo
  {journal} {Applied Physics Letters}\ }\textbf {\bibinfo {volume} {104}},\
  \bibinfo {eid} {022602} (\bibinfo {year} {2014})}\BibitemShut {NoStop}%
\bibitem [{\citenamefont {Petrovykh}\ \emph {et~al.}(1998)\citenamefont
  {Petrovykh}, \citenamefont {Altmann}, \citenamefont {H{\"o}chst},
  \citenamefont {Laubscher}, \citenamefont {Maat}, \citenamefont {Mankey},\
  and\ \citenamefont {Himpsel}}]{Petrovykh19983459}%
  \BibitemOpen
  \bibfield  {author} {\bibinfo {author} {\bibfnamefont {D.}~\bibnamefont
  {Petrovykh}}, \bibinfo {author} {\bibfnamefont {K.}~\bibnamefont {Altmann}},
  \bibinfo {author} {\bibfnamefont {H.}~\bibnamefont {H{\"o}chst}}, \bibinfo
  {author} {\bibfnamefont {M.}~\bibnamefont {Laubscher}}, \bibinfo {author}
  {\bibfnamefont {S.}~\bibnamefont {Maat}}, \bibinfo {author} {\bibfnamefont
  {G.}~\bibnamefont {Mankey}}, \ and\ \bibinfo {author} {\bibfnamefont
  {F.}~\bibnamefont {Himpsel}},\ }\href
  {http://www.scopus.com/inward/record.url?eid=2-s2.0-0001464491&partnerID=40&md5=7ce88f2f16f31b32616b1cf46245188c}
  {\bibfield  {journal} {\bibinfo  {journal} {Applied Physics Letters}\
  }\textbf {\bibinfo {volume} {73}},\ \bibinfo {pages} {3459} (\bibinfo {year}
  {1998})}\BibitemShut {NoStop}%
\bibitem [{\citenamefont {Villamor}\ \emph {et~al.}(2013)\citenamefont
  {Villamor}, \citenamefont {Isasa}, \citenamefont {Hueso},\ and\ \citenamefont
  {Casanova}}]{PhysRevB.88.184411}%
  \BibitemOpen
  \bibfield  {author} {\bibinfo {author} {\bibfnamefont {E.}~\bibnamefont
  {Villamor}}, \bibinfo {author} {\bibfnamefont {M.}~\bibnamefont {Isasa}},
  \bibinfo {author} {\bibfnamefont {L.~E.}\ \bibnamefont {Hueso}}, \ and\
  \bibinfo {author} {\bibfnamefont {F.}~\bibnamefont {Casanova}},\ }\href
  {\doibase 10.1103/PhysRevB.88.184411} {\bibfield  {journal} {\bibinfo
  {journal} {Phys. Rev. B}\ }\textbf {\bibinfo {volume} {88}},\ \bibinfo
  {pages} {184411} (\bibinfo {year} {2013})}\BibitemShut {NoStop}%
\bibitem [{\citenamefont {Haidar}\ and\ \citenamefont
  {Bailleul}(2013)}]{Haidar_2013}%
  \BibitemOpen
  \bibfield  {author} {\bibinfo {author} {\bibfnamefont {M.}~\bibnamefont
  {Haidar}}\ and\ \bibinfo {author} {\bibfnamefont {M.}~\bibnamefont
  {Bailleul}},\ }\href@noop {} {\bibfield  {journal} {\bibinfo  {journal}
  {Physical Review B}\ }\textbf {\bibinfo {volume} {88}},\ \bibinfo {pages}
  {054417} (\bibinfo {year} {2013})}\BibitemShut {NoStop}%
\bibitem [{\citenamefont {Slater}(1937)}]{Slater1937385}%
  \BibitemOpen
  \bibfield  {author} {\bibinfo {author} {\bibfnamefont {J.}~\bibnamefont
  {Slater}},\ }\href
  {http://www.scopus.com/inward/record.url?eid=2-s2.0-0000707809&partnerID=40&md5=eaa0944fca51f28d26d8b5b1bf2b38f7}
  {\bibfield  {journal} {\bibinfo  {journal} {Journal of Applied Physics}\
  }\textbf {\bibinfo {volume} {8}},\ \bibinfo {pages} {385} (\bibinfo {year}
  {1937})}\BibitemShut {NoStop}%
\bibitem [{\citenamefont {Collins}\ and\ \citenamefont
  {Wheeler}(1963)}]{Collins1963633}%
  \BibitemOpen
  \bibfield  {author} {\bibinfo {author} {\bibfnamefont {M.}~\bibnamefont
  {Collins}}\ and\ \bibinfo {author} {\bibfnamefont {D.}~\bibnamefont
  {Wheeler}},\ }\href
  {http://www.scopus.com/inward/record.url?eid=2-s2.0-0007029661&partnerID=40&md5=15bf39e99d1d88432c36d2c8f81369ee}
  {\bibfield  {journal} {\bibinfo  {journal} {Proceedings of the Physical
  Society}\ }\textbf {\bibinfo {volume} {82}},\ \bibinfo {pages} {633}
  (\bibinfo {year} {1963})}\BibitemShut {NoStop}%
\bibitem [{\citenamefont {Getzlaff}(2008)}]{Getzlaff20081}%
  \BibitemOpen
  \bibfield  {author} {\bibinfo {author} {\bibfnamefont {M.}~\bibnamefont
  {Getzlaff}},\ }\href
  {http://www.scopus.com/inward/record.url?eid=2-s2.0-84892242251&partnerID=40&md5=2759231e6116f9c1d656b7468ea244b3}
  {\emph {\bibinfo {title} {Fundamentals of magnetism}}}\ (\bibinfo
  {publisher} {Springer},\ \bibinfo {year} {2008})\BibitemShut {NoStop}%
\bibitem [{\citenamefont {Kakehashi}(2012)}]{Kakehashi}%
  \BibitemOpen
  \bibfield  {author} {\bibinfo {author} {\bibfnamefont {Y.}~\bibnamefont
  {Kakehashi}},\ }\href@noop {} {\emph {\bibinfo {title} {Modern Theory of
  Magnetism in Metals and Alloys}}}\ (\bibinfo  {publisher} {Springer},\
  \bibinfo {year} {2012})\BibitemShut {NoStop}%
\bibitem [{\citenamefont {Kittel}(1976)}]{Kittel_book}%
  \BibitemOpen
  \bibfield  {author} {\bibinfo {author} {\bibfnamefont {C.}~\bibnamefont
  {Kittel}},\ }\href@noop {} {\emph {\bibinfo {title} {Introduction to Solid
  State Physics}}}\ (\bibinfo  {publisher} {Wiley, New York},\ \bibinfo {year}
  {1976})\BibitemShut {NoStop}%
\bibitem [{\citenamefont {O'Handley}(2000)}]{OHandley_book}%
  \BibitemOpen
  \bibfield  {author} {\bibinfo {author} {\bibfnamefont {R.}~\bibnamefont
  {O'Handley}},\ }\href@noop {} {\emph {\bibinfo {title} {Modern magnetic
  materials}}}\ (\bibinfo  {publisher} {John Wiley and Sons, New York},\
  \bibinfo {year} {2000})\BibitemShut {NoStop}%
\bibitem [{\citenamefont {Jiles}(1998)}]{Jiles_Book}%
  \BibitemOpen
  \bibfield  {author} {\bibinfo {author} {\bibfnamefont {D.}~\bibnamefont
  {Jiles}},\ }\href@noop {} {\emph {\bibinfo {title} {Introduction to Magnetism
  and Magnetic materials}}}\ (\bibinfo  {publisher} {CRC press, Boca Raton},\
  \bibinfo {year} {1998})\BibitemShut {NoStop}%
\bibitem [{\citenamefont {Berkowitz}\ and\ \citenamefont
  {Kneller}(1969)}]{Berkowitz_Book}%
  \BibitemOpen
  \bibfield  {author} {\bibinfo {author} {\bibfnamefont {A.~E.}\ \bibnamefont
  {Berkowitz}}\ and\ \bibinfo {author} {\bibfnamefont {E.}~\bibnamefont
  {Kneller}},\ }\href@noop {} {\emph {\bibinfo {title} {Magnetism and
  Metallurgy}}}\ (\bibinfo  {publisher} {Academic press, New York},\ \bibinfo
  {year} {1969})\BibitemShut {NoStop}%
\bibitem [{\citenamefont {Craik}(1995)}]{Craik_book}%
  \BibitemOpen
  \bibfield  {author} {\bibinfo {author} {\bibfnamefont {D.~J.}\ \bibnamefont
  {Craik}},\ }\href@noop {} {\emph {\bibinfo {title} {Magnetism: Principles and
  Applications}}}\ (\bibinfo  {publisher} {Wiley, New York},\ \bibinfo {year}
  {1995})\BibitemShut {NoStop}%
\bibitem [{\citenamefont {Cullity}\ and\ \citenamefont
  {Graham}(2009)}]{Cullity_Book}%
  \BibitemOpen
  \bibfield  {author} {\bibinfo {author} {\bibfnamefont {B.~D.}\ \bibnamefont
  {Cullity}}\ and\ \bibinfo {author} {\bibfnamefont {C.~D.}\ \bibnamefont
  {Graham}},\ }\href@noop {} {\emph {\bibinfo {title} {Introduction to Magnetic
  Materials}}}\ (\bibinfo  {publisher} {Wiley, New Jersey},\ \bibinfo {year}
  {2009})\BibitemShut {NoStop}%
\bibitem [{\citenamefont {du~Tr{\'e}molet~de
  Lacheisserie}(1999)}]{Tremolet_Book}%
  \BibitemOpen
  \bibinfo {editor} {\bibfnamefont {E.}~\bibnamefont {du~Tr{\'e}molet~de
  Lacheisserie}},\ ed.,\ \href@noop {} {\emph {\bibinfo {title} {Magnetism I,
  Fundamentals}}}\ (\bibinfo  {publisher} {Springer},\ \bibinfo {year}
  {1999})\BibitemShut {NoStop}%
\bibitem [{\citenamefont {Vonsovskii}(1974)}]{Vonsovskii_book}%
  \BibitemOpen
  \bibfield  {author} {\bibinfo {author} {\bibfnamefont {S.~V.}\ \bibnamefont
  {Vonsovskii}},\ }\href@noop {} {\emph {\bibinfo {title} {Magnetism}}}\
  (\bibinfo  {publisher} {J. Wiley},\ \bibinfo {year} {1974})\BibitemShut
  {NoStop}%
\bibitem [{\citenamefont {Skomski}\ and\ \citenamefont
  {Coey}(1999)}]{Skomski_Book}%
  \BibitemOpen
  \bibfield  {author} {\bibinfo {author} {\bibfnamefont {R.}~\bibnamefont
  {Skomski}}\ and\ \bibinfo {author} {\bibfnamefont {J.}~\bibnamefont {Coey}},\
  }\href@noop {} {\emph {\bibinfo {title} {Permanent Magnetism}}}\ (\bibinfo
  {publisher} {CRC press},\ \bibinfo {year} {1999})\BibitemShut {NoStop}%
\bibitem [{\citenamefont {Bozorth}(1993)}]{Bozorth}%
  \BibitemOpen
  \bibfield  {author} {\bibinfo {author} {\bibfnamefont {R.~M.}\ \bibnamefont
  {Bozorth}},\ }\href@noop {} {\emph {\bibinfo {title} {Ferromagnetism}}}\
  (\bibinfo  {publisher} {Wiley-IEEE Press},\ \bibinfo {year}
  {1993})\BibitemShut {NoStop}%
\bibitem [{\citenamefont {Cable}\ and\ \citenamefont
  {Hicks}(1970)}]{Cable1970176}%
  \BibitemOpen
  \bibfield  {author} {\bibinfo {author} {\bibfnamefont {J.}~\bibnamefont
  {Cable}}\ and\ \bibinfo {author} {\bibfnamefont {T.}~\bibnamefont {Hicks}},\
  }\href
  {http://www.scopus.com/inward/record.url?eid=2-s2.0-0014813587&partnerID=40&md5=b98694183c9a413cccd18991b299e051}
  {\bibfield  {journal} {\bibinfo  {journal} {Physical Review B}\ }\textbf
  {\bibinfo {volume} {2}},\ \bibinfo {pages} {176} (\bibinfo {year}
  {1970})}\BibitemShut {NoStop}%
\bibitem [{\citenamefont {Dederichs}\ \emph {et~al.}(1991)\citenamefont
  {Dederichs}, \citenamefont {Zeller}, \citenamefont {Akai},\ and\
  \citenamefont {Ebert}}]{Dederichs1991241}%
  \BibitemOpen
  \bibfield  {author} {\bibinfo {author} {\bibfnamefont {P.}~\bibnamefont
  {Dederichs}}, \bibinfo {author} {\bibfnamefont {R.}~\bibnamefont {Zeller}},
  \bibinfo {author} {\bibfnamefont {H.}~\bibnamefont {Akai}}, \ and\ \bibinfo
  {author} {\bibfnamefont {H.}~\bibnamefont {Ebert}},\ }\href
  {http://www.scopus.com/inward/record.url?eid=2-s2.0-0026257767&partnerID=40&md5=bc201b4d6099c0808b0c19f251b39e7b}
  {\bibfield  {journal} {\bibinfo  {journal} {Journal of Magnetism and Magnetic
  Materials}\ }\textbf {\bibinfo {volume} {100}},\ \bibinfo {pages} {241}
  (\bibinfo {year} {1991})}\BibitemShut {NoStop}%
\bibitem [{\citenamefont {Turek}\ \emph {et~al.}(1994)\citenamefont {Turek},
  \citenamefont {Kudrnovsk\'y}, \citenamefont {Drchal},\ and\ \citenamefont
  {Weinberger}}]{PhysRevB.49.3352}%
  \BibitemOpen
  \bibfield  {author} {\bibinfo {author} {\bibfnamefont {I.}~\bibnamefont
  {Turek}}, \bibinfo {author} {\bibfnamefont {J.}~\bibnamefont {Kudrnovsk\'y}},
  \bibinfo {author} {\bibfnamefont {V.}~\bibnamefont {Drchal}}, \ and\ \bibinfo
  {author} {\bibfnamefont {P.}~\bibnamefont {Weinberger}},\ }\href {\doibase
  10.1103/PhysRevB.49.3352} {\bibfield  {journal} {\bibinfo  {journal} {Phys.
  Rev. B}\ }\textbf {\bibinfo {volume} {49}},\ \bibinfo {pages} {3352}
  (\bibinfo {year} {1994})}\BibitemShut {NoStop}%
\bibitem [{\citenamefont {Jacobs}\ \emph {et~al.}(1985)\citenamefont {Jacobs},
  \citenamefont {Babic},\ and\ \citenamefont {Xanthakis}}]{Jacobs19851941}%
  \BibitemOpen
  \bibfield  {author} {\bibinfo {author} {\bibfnamefont {R.}~\bibnamefont
  {Jacobs}}, \bibinfo {author} {\bibfnamefont {E.}~\bibnamefont {Babic}}, \
  and\ \bibinfo {author} {\bibfnamefont {J.}~\bibnamefont {Xanthakis}},\ }\href
  {http://www.scopus.com/inward/record.url?eid=2-s2.0-36149047073&partnerID=40&md5=138ffe6260bc868383f7e9b61b3a844a}
  {\bibfield  {journal} {\bibinfo  {journal} {Journal of Physics F: Metal
  Physics}\ }\textbf {\bibinfo {volume} {15}},\ \bibinfo {pages} {1941}
  (\bibinfo {year} {1985})}\BibitemShut {NoStop}%
\bibitem [{\citenamefont {Vernes}\ \emph {et~al.}(2003)\citenamefont {Vernes},
  \citenamefont {Ebert},\ and\ \citenamefont {Banhart}}]{Vernes2003}%
  \BibitemOpen
  \bibfield  {author} {\bibinfo {author} {\bibfnamefont {A.}~\bibnamefont
  {Vernes}}, \bibinfo {author} {\bibfnamefont {H.}~\bibnamefont {Ebert}}, \
  and\ \bibinfo {author} {\bibfnamefont {J.}~\bibnamefont {Banhart}},\ }\href
  {\doibase 10.1103/PhysRevB.68.134404} {\bibfield  {journal} {\bibinfo
  {journal} {Physical Review B}\ }\textbf {\bibinfo {volume} {68}},\ \bibinfo
  {pages} {134404} (\bibinfo {year} {2003})}\BibitemShut {NoStop}%
\bibitem [{\citenamefont {James}\ \emph {et~al.}(1999)\citenamefont {James},
  \citenamefont {Eriksson}, \citenamefont {Johansson},\ and\ \citenamefont
  {Abrikosov}}]{PhysRevB.59.419}%
  \BibitemOpen
  \bibfield  {author} {\bibinfo {author} {\bibfnamefont {P.}~\bibnamefont
  {James}}, \bibinfo {author} {\bibfnamefont {O.}~\bibnamefont {Eriksson}},
  \bibinfo {author} {\bibfnamefont {B.}~\bibnamefont {Johansson}}, \ and\
  \bibinfo {author} {\bibfnamefont {I.~A.}\ \bibnamefont {Abrikosov}},\ }\href
  {\doibase 10.1103/PhysRevB.59.419} {\bibfield  {journal} {\bibinfo  {journal}
  {Phys. Rev. B}\ }\textbf {\bibinfo {volume} {59}},\ \bibinfo {pages} {419}
  (\bibinfo {year} {1999})}\BibitemShut {NoStop}%
\bibitem [{\citenamefont {Choy}\ \emph {et~al.}(1999)\citenamefont {Choy},
  \citenamefont {Chen},\ and\ \citenamefont {Hershfield}}]{Choy_aip}%
  \BibitemOpen
  \bibfield  {author} {\bibinfo {author} {\bibfnamefont {T.-S.}\ \bibnamefont
  {Choy}}, \bibinfo {author} {\bibfnamefont {J.}~\bibnamefont {Chen}}, \ and\
  \bibinfo {author} {\bibfnamefont {S.}~\bibnamefont {Hershfield}},\
  }\href@noop {} {\bibfield  {journal} {\bibinfo  {journal} {Journal of Applied
  Physics}\ }\textbf {\bibinfo {volume} {86}},\ \bibinfo {pages} {562}
  (\bibinfo {year} {1999})}\BibitemShut {NoStop}%
\bibitem [{\citenamefont {Mijnarends}\ \emph {et~al.}(2002)\citenamefont
  {Mijnarends}, \citenamefont {Sahrakorpi}, \citenamefont {Lindroos},\ and\
  \citenamefont {Bansil}}]{PhysRevB.65.075106}%
  \BibitemOpen
  \bibfield  {author} {\bibinfo {author} {\bibfnamefont {P.~E.}\ \bibnamefont
  {Mijnarends}}, \bibinfo {author} {\bibfnamefont {S.}~\bibnamefont
  {Sahrakorpi}}, \bibinfo {author} {\bibfnamefont {M.}~\bibnamefont
  {Lindroos}}, \ and\ \bibinfo {author} {\bibfnamefont {A.}~\bibnamefont
  {Bansil}},\ }\href {\doibase 10.1103/PhysRevB.65.075106} {\bibfield
  {journal} {\bibinfo  {journal} {Phys. Rev. B}\ }\textbf {\bibinfo {volume}
  {65}},\ \bibinfo {pages} {075106} (\bibinfo {year} {2002})}\BibitemShut
  {NoStop}%
\bibitem [{\citenamefont {Smirnova}\ \emph {et~al.}(1999)\citenamefont
  {Smirnova}, \citenamefont {Abrikosov}, \citenamefont {Johansson},
  \citenamefont {Vekilov}, \citenamefont {Baranov}, \citenamefont {Stepanyuk},
  \citenamefont {Hergert},\ and\ \citenamefont
  {Dederichs}}]{Smirnova199914417}%
  \BibitemOpen
  \bibfield  {author} {\bibinfo {author} {\bibfnamefont {E.}~\bibnamefont
  {Smirnova}}, \bibinfo {author} {\bibfnamefont {I.}~\bibnamefont {Abrikosov}},
  \bibinfo {author} {\bibfnamefont {B.}~\bibnamefont {Johansson}}, \bibinfo
  {author} {\bibfnamefont {Y.}~\bibnamefont {Vekilov}}, \bibinfo {author}
  {\bibfnamefont {A.}~\bibnamefont {Baranov}}, \bibinfo {author} {\bibfnamefont
  {V.}~\bibnamefont {Stepanyuk}}, \bibinfo {author} {\bibfnamefont
  {W.}~\bibnamefont {Hergert}}, \ and\ \bibinfo {author} {\bibfnamefont
  {P.}~\bibnamefont {Dederichs}},\ }\href
  {http://www.scopus.com/inward/record.url?eid=2-s2.0-0001020840&partnerID=40&md5=051f9a409d7405ea6284957a3c3ead14}
  {\bibfield  {journal} {\bibinfo  {journal} {Physical Review B - Condensed
  Matter and Materials Physics}\ }\textbf {\bibinfo {volume} {59}},\ \bibinfo
  {pages} {14417} (\bibinfo {year} {1999})}\BibitemShut {NoStop}%
\bibitem [{\citenamefont {Paduani}\ and\ \citenamefont
  {Krause}(1999)}]{Paduani_JAP_1999}%
  \BibitemOpen
  \bibfield  {author} {\bibinfo {author} {\bibfnamefont {C.}~\bibnamefont
  {Paduani}}\ and\ \bibinfo {author} {\bibfnamefont {J.~C.}\ \bibnamefont
  {Krause}},\ }\href {\doibase http://dx.doi.org/10.1063/1.370769} {\bibfield
  {journal} {\bibinfo  {journal} {Journal of Applied Physics}\ }\textbf
  {\bibinfo {volume} {86}},\ \bibinfo {pages} {578} (\bibinfo {year}
  {1999})}\BibitemShut {NoStop}%
\bibitem [{\citenamefont {Serena}\ and\ \citenamefont
  {Garcia}(1994)}]{PhysRevB.50.944}%
  \BibitemOpen
  \bibfield  {author} {\bibinfo {author} {\bibfnamefont {P.~A.}\ \bibnamefont
  {Serena}}\ and\ \bibinfo {author} {\bibfnamefont {N.}~\bibnamefont
  {Garcia}},\ }\href {\doibase 10.1103/PhysRevB.50.944} {\bibfield  {journal}
  {\bibinfo  {journal} {Phys. Rev. B}\ }\textbf {\bibinfo {volume} {50}},\
  \bibinfo {pages} {944} (\bibinfo {year} {1994})}\BibitemShut {NoStop}%
\bibitem [{\citenamefont {Zeller}(1987)}]{Zeller19872123}%
  \BibitemOpen
  \bibfield  {author} {\bibinfo {author} {\bibfnamefont {R.}~\bibnamefont
  {Zeller}},\ }\href
  {http://www.scopus.com/inward/record.url?eid=2-s2.0-0023433620&partnerID=40&md5=1674433164f162aa6689ae2ee86119f4}
  {\bibfield  {journal} {\bibinfo  {journal} {Journal of Physics F: Metal
  Physics}\ }\textbf {\bibinfo {volume} {17}},\ \bibinfo {pages} {2123}
  (\bibinfo {year} {1987})}\BibitemShut {NoStop}%
\bibitem [{\citenamefont {Stepanyuk}\ \emph {et~al.}(1994)\citenamefont
  {Stepanyuk}, \citenamefont {Zeller}, \citenamefont {Dederichs},\ and\
  \citenamefont {Mertig}}]{PhysRevB.49.5157}%
  \BibitemOpen
  \bibfield  {author} {\bibinfo {author} {\bibfnamefont {V.~S.}\ \bibnamefont
  {Stepanyuk}}, \bibinfo {author} {\bibfnamefont {R.}~\bibnamefont {Zeller}},
  \bibinfo {author} {\bibfnamefont {P.~H.}\ \bibnamefont {Dederichs}}, \ and\
  \bibinfo {author} {\bibfnamefont {I.}~\bibnamefont {Mertig}},\ }\href
  {\doibase 10.1103/PhysRevB.49.5157} {\bibfield  {journal} {\bibinfo
  {journal} {Phys. Rev. B}\ }\textbf {\bibinfo {volume} {49}},\ \bibinfo
  {pages} {5157} (\bibinfo {year} {1994})}\BibitemShut {NoStop}%
\bibitem [{\citenamefont {Yu}\ \emph {et~al.}(2008)\citenamefont {Yu},
  \citenamefont {Jin}, \citenamefont {Kudrnovsk\'y}, \citenamefont {Wang},\
  and\ \citenamefont {Bruno}}]{PhysRevB.77.054431}%
  \BibitemOpen
  \bibfield  {author} {\bibinfo {author} {\bibfnamefont {P.}~\bibnamefont
  {Yu}}, \bibinfo {author} {\bibfnamefont {X.~F.}\ \bibnamefont {Jin}},
  \bibinfo {author} {\bibfnamefont {J.}~\bibnamefont {Kudrnovsk\'y}}, \bibinfo
  {author} {\bibfnamefont {D.~S.}\ \bibnamefont {Wang}}, \ and\ \bibinfo
  {author} {\bibfnamefont {P.}~\bibnamefont {Bruno}},\ }\href {\doibase
  10.1103/PhysRevB.77.054431} {\bibfield  {journal} {\bibinfo  {journal} {Phys.
  Rev. B}\ }\textbf {\bibinfo {volume} {77}},\ \bibinfo {pages} {054431}
  (\bibinfo {year} {2008})}\BibitemShut {NoStop}%
\bibitem [{\citenamefont {Moghadam}\ and\ \citenamefont
  {Stocks}(2005)}]{PhysRevB.71.134421}%
  \BibitemOpen
  \bibfield  {author} {\bibinfo {author} {\bibfnamefont {N.~Y.}\ \bibnamefont
  {Moghadam}}\ and\ \bibinfo {author} {\bibfnamefont {G.~M.}\ \bibnamefont
  {Stocks}},\ }\href {\doibase 10.1103/PhysRevB.71.134421} {\bibfield
  {journal} {\bibinfo  {journal} {Phys. Rev. B}\ }\textbf {\bibinfo {volume}
  {71}},\ \bibinfo {pages} {134421} (\bibinfo {year} {2005})}\BibitemShut
  {NoStop}%
\bibitem [{\citenamefont {Minar}\ \emph {et~al.}(2014)\citenamefont {Minar},
  \citenamefont {Mankovsky}, \citenamefont {Sipr}, \citenamefont {Benea},\ and\
  \citenamefont {Ebert}}]{Minar_JPhys_2014}%
  \BibitemOpen
  \bibfield  {author} {\bibinfo {author} {\bibfnamefont {J.}~\bibnamefont
  {Minar}}, \bibinfo {author} {\bibfnamefont {S.}~\bibnamefont {Mankovsky}},
  \bibinfo {author} {\bibnamefont {Sipr}}, \bibinfo {author} {\bibfnamefont
  {D.}~\bibnamefont {Benea}}, \ and\ \bibinfo {author} {\bibfnamefont
  {H.}~\bibnamefont {Ebert}},\ }\href
  {http://stacks.iop.org/0953-8984/26/i=27/a=274206} {\bibfield  {journal}
  {\bibinfo  {journal} {Journal of Physics: Condensed Matter}\ }\textbf
  {\bibinfo {volume} {26}},\ \bibinfo {pages} {274206} (\bibinfo {year}
  {2014})}\BibitemShut {NoStop}%
\bibitem [{\citenamefont {Mathias}\ \emph {et~al.}(2012)\citenamefont
  {Mathias}, \citenamefont {La-O-Vorakiat}, \citenamefont {Grychtol},
  \citenamefont {Granitzka}, \citenamefont {Turgut}, \citenamefont {Shaw},
  \citenamefont {Adam}, \citenamefont {Nembach}, \citenamefont {Siemens},
  \citenamefont {Eich}, \citenamefont {Schneider}, \citenamefont {Silva},
  \citenamefont {Aeschlimann}, \citenamefont {Murnane},\ and\ \citenamefont
  {Kapteyn}}]{Mathias20124792}%
  \BibitemOpen
  \bibfield  {author} {\bibinfo {author} {\bibfnamefont {S.}~\bibnamefont
  {Mathias}}, \bibinfo {author} {\bibfnamefont {C.}~\bibnamefont
  {La-O-Vorakiat}}, \bibinfo {author} {\bibfnamefont {P.}~\bibnamefont
  {Grychtol}}, \bibinfo {author} {\bibfnamefont {P.}~\bibnamefont {Granitzka}},
  \bibinfo {author} {\bibfnamefont {E.}~\bibnamefont {Turgut}}, \bibinfo
  {author} {\bibfnamefont {J.}~\bibnamefont {Shaw}}, \bibinfo {author}
  {\bibfnamefont {R.}~\bibnamefont {Adam}}, \bibinfo {author} {\bibfnamefont
  {H.}~\bibnamefont {Nembach}}, \bibinfo {author} {\bibfnamefont
  {M.}~\bibnamefont {Siemens}}, \bibinfo {author} {\bibfnamefont
  {S.}~\bibnamefont {Eich}}, \bibinfo {author} {\bibfnamefont {C.}~\bibnamefont
  {Schneider}}, \bibinfo {author} {\bibfnamefont {T.}~\bibnamefont {Silva}},
  \bibinfo {author} {\bibfnamefont {M.}~\bibnamefont {Aeschlimann}}, \bibinfo
  {author} {\bibfnamefont {M.}~\bibnamefont {Murnane}}, \ and\ \bibinfo
  {author} {\bibfnamefont {H.}~\bibnamefont {Kapteyn}},\ }\href
  {http://www.scopus.com/inward/record.url?eid=2-s2.0-84859476532&partnerID=40&md5=7efff6aad19cf6df7a261df4ebd7f420}
  {\bibfield  {journal} {\bibinfo  {journal} {Proceedings of the National
  Academy of Sciences of the United States of America}\ }\textbf {\bibinfo
  {volume} {109}},\ \bibinfo {pages} {4792} (\bibinfo {year}
  {2012})}\BibitemShut {NoStop}%
\bibitem [{\citenamefont {G\"unther}\ \emph {et~al.}(2014)\citenamefont
  {G\"unther}, \citenamefont {Spezzani}, \citenamefont {Ciprian}, \citenamefont
  {Grazioli}, \citenamefont {Ressel}, \citenamefont {Coreno}, \citenamefont
  {Poletto}, \citenamefont {Miotti}, \citenamefont {Sacchi}, \citenamefont
  {Panaccione}, \citenamefont {Uhl\'{i}\ifmmode~\check{r}\else \v{r}\fi{}},
  \citenamefont {Fullerton}, \citenamefont {De~Ninno},\ and\ \citenamefont
  {Back}}]{PhysRevB.90.180407}%
  \BibitemOpen
  \bibfield  {author} {\bibinfo {author} {\bibfnamefont {S.}~\bibnamefont
  {G\"unther}}, \bibinfo {author} {\bibfnamefont {C.}~\bibnamefont {Spezzani}},
  \bibinfo {author} {\bibfnamefont {R.}~\bibnamefont {Ciprian}}, \bibinfo
  {author} {\bibfnamefont {C.}~\bibnamefont {Grazioli}}, \bibinfo {author}
  {\bibfnamefont {B.}~\bibnamefont {Ressel}}, \bibinfo {author} {\bibfnamefont
  {M.}~\bibnamefont {Coreno}}, \bibinfo {author} {\bibfnamefont
  {L.}~\bibnamefont {Poletto}}, \bibinfo {author} {\bibfnamefont
  {P.}~\bibnamefont {Miotti}}, \bibinfo {author} {\bibfnamefont
  {M.}~\bibnamefont {Sacchi}}, \bibinfo {author} {\bibfnamefont
  {G.}~\bibnamefont {Panaccione}}, \bibinfo {author} {\bibfnamefont
  {V.}~\bibnamefont {Uhl\'{i}\ifmmode~\check{r}\else \v{r}\fi{}}}, \bibinfo
  {author} {\bibfnamefont {E.~E.}\ \bibnamefont {Fullerton}}, \bibinfo {author}
  {\bibfnamefont {G.}~\bibnamefont {De~Ninno}}, \ and\ \bibinfo {author}
  {\bibfnamefont {C.~H.}\ \bibnamefont {Back}},\ }\href {\doibase
  10.1103/PhysRevB.90.180407} {\bibfield  {journal} {\bibinfo  {journal} {Phys.
  Rev. B}\ }\textbf {\bibinfo {volume} {90}},\ \bibinfo {pages} {180407}
  (\bibinfo {year} {2014})},\ \bibinfo {note} {see Supplemental}\BibitemShut
  {NoStop}%
\bibitem [{Som()}]{Somnath_JANA}%
  \BibitemOpen
  \href@noop {} {\bibinfo  {journal} {Private communication: Somnath Jana,
  Joachim Terschluesen, Robert Stefanuik, Johan S{\"o}derstr{\"o}m, and Olof
  Karis (Unpublished, Uppsala University, Sweden)}\ }\BibitemShut {NoStop}%
\bibitem [{\citenamefont {Eschenlohr}(2012)}]{EschenlohrThesis}%
  \BibitemOpen
\bibfield  {journal} {  }\bibfield  {author} {\bibinfo {author} {\bibfnamefont
  {A.}~\bibnamefont {Eschenlohr}},\ }\href@noop {} {Ph.D. thesis},\ \bibinfo
  {school} {Universit\"at Potsdam} (\bibinfo {year} {2012})\BibitemShut
  {NoStop}%
\bibitem [{\citenamefont {Koopmans}\ \emph {et~al.}(2010)\citenamefont
  {Koopmans}, \citenamefont {Malinowski}, \citenamefont {Dalla~Longa},
  \citenamefont {Steiauf}, \citenamefont {Fahnle}, \citenamefont {Roth},
  \citenamefont {Cinchetti},\ and\ \citenamefont
  {Aeschlimann}}]{Koopmans_NatMat_2010}%
  \BibitemOpen
  \bibfield  {author} {\bibinfo {author} {\bibfnamefont {B.}~\bibnamefont
  {Koopmans}}, \bibinfo {author} {\bibfnamefont {G.}~\bibnamefont
  {Malinowski}}, \bibinfo {author} {\bibfnamefont {F.}~\bibnamefont
  {Dalla~Longa}}, \bibinfo {author} {\bibfnamefont {D.}~\bibnamefont
  {Steiauf}}, \bibinfo {author} {\bibfnamefont {M.}~\bibnamefont {Fahnle}},
  \bibinfo {author} {\bibfnamefont {T.}~\bibnamefont {Roth}}, \bibinfo {author}
  {\bibfnamefont {M.}~\bibnamefont {Cinchetti}}, \ and\ \bibinfo {author}
  {\bibfnamefont {M.}~\bibnamefont {Aeschlimann}},\ }\href
  {http://dx.doi.org/10.1038/nmat2593} {\bibfield  {journal} {\bibinfo
  {journal} {Nat Mater}\ }\textbf {\bibinfo {volume} {9}},\ \bibinfo {pages}
  {259} (\bibinfo {year} {2010})}\BibitemShut {NoStop}%
\bibitem [{\citenamefont {Carva}\ \emph {et~al.}(2013)\citenamefont {Carva},
  \citenamefont {Battiato}, \citenamefont {Legut},\ and\ \citenamefont
  {Oppeneer}}]{Carva_PRB_2013}%
  \BibitemOpen
  \bibfield  {author} {\bibinfo {author} {\bibfnamefont {K.}~\bibnamefont
  {Carva}}, \bibinfo {author} {\bibfnamefont {M.}~\bibnamefont {Battiato}},
  \bibinfo {author} {\bibfnamefont {D.}~\bibnamefont {Legut}}, \ and\ \bibinfo
  {author} {\bibfnamefont {P.~M.}\ \bibnamefont {Oppeneer}},\ }\href@noop {}
  {\bibfield  {journal} {\bibinfo  {journal} {Phys. Rev. B}\ }\textbf {\bibinfo
  {volume} {87}},\ \bibinfo {pages} {184425} (\bibinfo {year}
  {2013})}\BibitemShut {NoStop}%
\bibitem [{\citenamefont {Nakajima}\ \emph {et~al.}(1999)\citenamefont
  {Nakajima}, \citenamefont {St\"ohr},\ and\ \citenamefont
  {Idzerda}}]{PhysRevB.59.6421}%
  \BibitemOpen
  \bibfield  {author} {\bibinfo {author} {\bibfnamefont {R.}~\bibnamefont
  {Nakajima}}, \bibinfo {author} {\bibfnamefont {J.}~\bibnamefont {St\"ohr}}, \
  and\ \bibinfo {author} {\bibfnamefont {Y.~U.}\ \bibnamefont {Idzerda}},\
  }\href {\doibase 10.1103/PhysRevB.59.6421} {\bibfield  {journal} {\bibinfo
  {journal} {Phys. Rev. B}\ }\textbf {\bibinfo {volume} {59}},\ \bibinfo
  {pages} {6421} (\bibinfo {year} {1999})}\BibitemShut {NoStop}%
\bibitem [{\citenamefont {Ebert}\ \emph {et~al.}(2011)\citenamefont {Ebert},
  \citenamefont {K{\"o}dderitzsch},\ and\ \citenamefont
  {Min{\'a}r}}]{0034-4885-74-9-096501}%
  \BibitemOpen
  \bibfield  {author} {\bibinfo {author} {\bibfnamefont {H.}~\bibnamefont
  {Ebert}}, \bibinfo {author} {\bibfnamefont {D.}~\bibnamefont
  {K{\"o}dderitzsch}}, \ and\ \bibinfo {author} {\bibfnamefont
  {J.}~\bibnamefont {Min{\'a}r}},\ }\href
  {http://stacks.iop.org/0034-4885/74/i=9/a=096501} {\bibfield  {journal}
  {\bibinfo  {journal} {Reports on Progress in Physics}\ }\textbf {\bibinfo
  {volume} {74}},\ \bibinfo {pages} {096501} (\bibinfo {year}
  {2011})}\BibitemShut {NoStop}%
\bibitem [{SPR()}]{SPRKKR}%
  \BibitemOpen
  \href@noop {} {\emph {\bibinfo {title}
  {http://ebert.cup.uni-muenchen.de/SPRKKR}}}\BibitemShut {NoStop}%
\bibitem [{\citenamefont {Perdew}\ \emph {et~al.}(1996)\citenamefont {Perdew},
  \citenamefont {Burke},\ and\ \citenamefont
  {Ernzerhof}}]{PhysRevLett.77.3865}%
  \BibitemOpen
  \bibfield  {author} {\bibinfo {author} {\bibfnamefont {J.~P.}\ \bibnamefont
  {Perdew}}, \bibinfo {author} {\bibfnamefont {K.}~\bibnamefont {Burke}}, \
  and\ \bibinfo {author} {\bibfnamefont {M.}~\bibnamefont {Ernzerhof}},\ }\href
  {\doibase 10.1103/PhysRevLett.77.3865} {\bibfield  {journal} {\bibinfo
  {journal} {Phys. Rev. Lett.}\ }\textbf {\bibinfo {volume} {77}},\ \bibinfo
  {pages} {3865} (\bibinfo {year} {1996})}\BibitemShut {NoStop}%
\bibitem [{\citenamefont {Chen}\ \emph {et~al.}(1995)\citenamefont {Chen},
  \citenamefont {Idzerda}, \citenamefont {Lin}, \citenamefont {Smith},
  \citenamefont {Meigs}, \citenamefont {Chaban}, \citenamefont {Ho},
  \citenamefont {Pellegrin},\ and\ \citenamefont {Sette}}]{PhysRevLett.75.152}%
  \BibitemOpen
  \bibfield  {author} {\bibinfo {author} {\bibfnamefont {C.~T.}\ \bibnamefont
  {Chen}}, \bibinfo {author} {\bibfnamefont {Y.~U.}\ \bibnamefont {Idzerda}},
  \bibinfo {author} {\bibfnamefont {H.-J.}\ \bibnamefont {Lin}}, \bibinfo
  {author} {\bibfnamefont {N.~V.}\ \bibnamefont {Smith}}, \bibinfo {author}
  {\bibfnamefont {G.}~\bibnamefont {Meigs}}, \bibinfo {author} {\bibfnamefont
  {E.}~\bibnamefont {Chaban}}, \bibinfo {author} {\bibfnamefont {G.~H.}\
  \bibnamefont {Ho}}, \bibinfo {author} {\bibfnamefont {E.}~\bibnamefont
  {Pellegrin}}, \ and\ \bibinfo {author} {\bibfnamefont {F.}~\bibnamefont
  {Sette}},\ }\href {\doibase 10.1103/PhysRevLett.75.152} {\bibfield  {journal}
  {\bibinfo  {journal} {Phys. Rev. Lett.}\ }\textbf {\bibinfo {volume} {75}},\
  \bibinfo {pages} {152} (\bibinfo {year} {1995})}\BibitemShut {NoStop}%
\bibitem [{\citenamefont {Nguyen}\ \emph {et~al.}(2014)\citenamefont {Nguyen},
  \citenamefont {Knut}, \citenamefont {Fallahi}, \citenamefont {Chung},
  \citenamefont {Le}, \citenamefont {Mohseni}, \citenamefont {Karis},
  \citenamefont {Peredkov}, \citenamefont {Dumas}, \citenamefont {Miller},\
  and\ \citenamefont {\AA{}kerman}}]{PhysRevApplied.2.044014}%
  \BibitemOpen
  \bibfield  {author} {\bibinfo {author} {\bibfnamefont {T.~N.~A.}\
  \bibnamefont {Nguyen}}, \bibinfo {author} {\bibfnamefont {R.}~\bibnamefont
  {Knut}}, \bibinfo {author} {\bibfnamefont {V.}~\bibnamefont {Fallahi}},
  \bibinfo {author} {\bibfnamefont {S.}~\bibnamefont {Chung}}, \bibinfo
  {author} {\bibfnamefont {Q.~T.}\ \bibnamefont {Le}}, \bibinfo {author}
  {\bibfnamefont {S.~M.}\ \bibnamefont {Mohseni}}, \bibinfo {author}
  {\bibfnamefont {O.}~\bibnamefont {Karis}}, \bibinfo {author} {\bibfnamefont
  {S.}~\bibnamefont {Peredkov}}, \bibinfo {author} {\bibfnamefont {R.~K.}\
  \bibnamefont {Dumas}}, \bibinfo {author} {\bibfnamefont {C.~W.}\ \bibnamefont
  {Miller}}, \ and\ \bibinfo {author} {\bibfnamefont {J.}~\bibnamefont
  {\AA{}kerman}},\ }\href {\doibase 10.1103/PhysRevApplied.2.044014} {\bibfield
   {journal} {\bibinfo  {journal} {Phys. Rev. Applied}\ }\textbf {\bibinfo
  {volume} {2}},\ \bibinfo {pages} {044014} (\bibinfo {year} {2014})},\
  \bibinfo {note} {and references within.}\BibitemShut {Stop}%
\bibitem [{\citenamefont {Collins}\ and\ \citenamefont
  {Forsyth}(1963)}]{Collins_someTAlloys}%
  \BibitemOpen
  \bibfield  {author} {\bibinfo {author} {\bibfnamefont {M.~F.}\ \bibnamefont
  {Collins}}\ and\ \bibinfo {author} {\bibfnamefont {J.~B.}\ \bibnamefont
  {Forsyth}},\ }\href {\doibase 10.1080/14786436308211141} {\bibfield
  {journal} {\bibinfo  {journal} {Philosophical Magazine}\ }\textbf {\bibinfo
  {volume} {8}},\ \bibinfo {pages} {401} (\bibinfo {year} {1963})}\BibitemShut
  {NoStop}%
\bibitem [{\citenamefont {Medina}\ and\ \citenamefont
  {Cable}(1977)}]{PhysRevB.15.1539}%
  \BibitemOpen
  \bibfield  {author} {\bibinfo {author} {\bibfnamefont {R.~A.}\ \bibnamefont
  {Medina}}\ and\ \bibinfo {author} {\bibfnamefont {J.~W.}\ \bibnamefont
  {Cable}},\ }\href {\doibase 10.1103/PhysRevB.15.1539} {\bibfield  {journal}
  {\bibinfo  {journal} {Phys. Rev. B}\ }\textbf {\bibinfo {volume} {15}},\
  \bibinfo {pages} {1539} (\bibinfo {year} {1977})}\BibitemShut {NoStop}%
\bibitem [{\citenamefont {Aldred}\ \emph {et~al.}(1973)\citenamefont {Aldred},
  \citenamefont {Rainford}, \citenamefont {Hicks},\ and\ \citenamefont
  {Kouvel}}]{Aldred1973218}%
  \BibitemOpen
  \bibfield  {author} {\bibinfo {author} {\bibfnamefont {A.}~\bibnamefont
  {Aldred}}, \bibinfo {author} {\bibfnamefont {B.}~\bibnamefont {Rainford}},
  \bibinfo {author} {\bibfnamefont {T.}~\bibnamefont {Hicks}}, \ and\ \bibinfo
  {author} {\bibfnamefont {J.}~\bibnamefont {Kouvel}},\ }\href
  {http://www.scopus.com/inward/record.url?eid=2-s2.0-0038599889&partnerID=40&md5=13c6b1b273c3b6bc4dd23c117541f357}
  {\bibfield  {journal} {\bibinfo  {journal} {Physical Review B}\ }\textbf
  {\bibinfo {volume} {7}},\ \bibinfo {pages} {218} (\bibinfo {year}
  {1973})}\BibitemShut {NoStop}%
\bibitem [{\citenamefont {Tersoff}\ and\ \citenamefont
  {Falicov}(1982)}]{PhysRevB.25.4937}%
  \BibitemOpen
  \bibfield  {author} {\bibinfo {author} {\bibfnamefont {J.}~\bibnamefont
  {Tersoff}}\ and\ \bibinfo {author} {\bibfnamefont {L.~M.}\ \bibnamefont
  {Falicov}},\ }\href {\doibase 10.1103/PhysRevB.25.4937} {\bibfield  {journal}
  {\bibinfo  {journal} {Phys. Rev. B}\ }\textbf {\bibinfo {volume} {25}},\
  \bibinfo {pages} {4937} (\bibinfo {year} {1982})}\BibitemShut {NoStop}%
\bibitem [{\citenamefont {Vogel}\ and\ \citenamefont
  {Sacchi}(1996)}]{PhysRevB.53.3409}%
  \BibitemOpen
  \bibfield  {author} {\bibinfo {author} {\bibfnamefont {J.}~\bibnamefont
  {Vogel}}\ and\ \bibinfo {author} {\bibfnamefont {M.}~\bibnamefont {Sacchi}},\
  }\href {\doibase 10.1103/PhysRevB.53.3409} {\bibfield  {journal} {\bibinfo
  {journal} {Phys. Rev. B}\ }\textbf {\bibinfo {volume} {53}},\ \bibinfo
  {pages} {3409} (\bibinfo {year} {1996})}\BibitemShut {NoStop}%
\bibitem [{\citenamefont {Glaubitz}\ \emph {et~al.}(2011)\citenamefont
  {Glaubitz}, \citenamefont {Buschhorn}, \citenamefont {Br{\"u}ssing},
  \citenamefont {Abrudan},\ and\ \citenamefont {Zabel}}]{Glaubitz2011}%
  \BibitemOpen
  \bibfield  {author} {\bibinfo {author} {\bibfnamefont {B.}~\bibnamefont
  {Glaubitz}}, \bibinfo {author} {\bibfnamefont {S.}~\bibnamefont {Buschhorn}},
  \bibinfo {author} {\bibfnamefont {F.}~\bibnamefont {Br{\"u}ssing}}, \bibinfo
  {author} {\bibfnamefont {R.}~\bibnamefont {Abrudan}}, \ and\ \bibinfo
  {author} {\bibfnamefont {H.}~\bibnamefont {Zabel}},\ }\href
  {http://www.scopus.com/inward/record.url?eid=2-s2.0-79959922970&partnerID=40&md5=29a16f111c7e3e6be68b6f45c8226d31}
  {\bibfield  {journal} {\bibinfo  {journal} {Journal of Physics Condensed
  Matter}\ }\textbf {\bibinfo {volume} {23}} (\bibinfo {year}
  {2011})}\BibitemShut {NoStop}%
\bibitem [{\citenamefont {Stohr}(1999)}]{Stohr1999470}%
  \BibitemOpen
  \bibfield  {author} {\bibinfo {author} {\bibfnamefont {J.}~\bibnamefont
  {Stohr}},\ }\href
  {http://www.scopus.com/inward/record.url?eid=2-s2.0-0038237201&partnerID=40&md5=f8e797ee7b4fbe068a4e173b49dd5647}
  {\bibfield  {journal} {\bibinfo  {journal} {Journal of Magnetism and Magnetic
  Materials}\ }\textbf {\bibinfo {volume} {200}},\ \bibinfo {pages} {470}
  (\bibinfo {year} {1999})}\BibitemShut {NoStop}%
\bibitem [{\citenamefont {Meitzner}\ \emph {et~al.}(1992)\citenamefont
  {Meitzner}, \citenamefont {Fischer},\ and\ \citenamefont
  {Sinfelt}}]{Meitzner1992219}%
  \BibitemOpen
  \bibfield  {author} {\bibinfo {author} {\bibfnamefont {G.}~\bibnamefont
  {Meitzner}}, \bibinfo {author} {\bibfnamefont {D.}~\bibnamefont {Fischer}}, \
  and\ \bibinfo {author} {\bibfnamefont {J.}~\bibnamefont {Sinfelt}},\ }\href
  {http://www.scopus.com/inward/record.url?eid=2-s2.0-0000544851&partnerID=40&md5=db6a0e177fa9cf48780a3a93822ae386}
  {\bibfield  {journal} {\bibinfo  {journal} {Catalysis Letters}\ }\textbf
  {\bibinfo {volume} {15}},\ \bibinfo {pages} {219} (\bibinfo {year}
  {1992})}\BibitemShut {NoStop}%
\bibitem [{\citenamefont {Stefanou}\ \emph {et~al.}(1987)\citenamefont
  {Stefanou}, \citenamefont {Oswald}, \citenamefont {Zeller},\ and\
  \citenamefont {Dederichs}}]{PhysRevB.35.6911}%
  \BibitemOpen
  \bibfield  {author} {\bibinfo {author} {\bibfnamefont {N.}~\bibnamefont
  {Stefanou}}, \bibinfo {author} {\bibfnamefont {A.}~\bibnamefont {Oswald}},
  \bibinfo {author} {\bibfnamefont {R.}~\bibnamefont {Zeller}}, \ and\ \bibinfo
  {author} {\bibfnamefont {P.~H.}\ \bibnamefont {Dederichs}},\ }\href {\doibase
  10.1103/PhysRevB.35.6911} {\bibfield  {journal} {\bibinfo  {journal} {Phys.
  Rev. B}\ }\textbf {\bibinfo {volume} {35}},\ \bibinfo {pages} {6911}
  (\bibinfo {year} {1987})}\BibitemShut {NoStop}%
\bibitem [{\citenamefont {Campbell}\ and\ \citenamefont
  {Gom{\`e}s}(1967)}]{Campbell1967319}%
  \BibitemOpen
  \bibfield  {author} {\bibinfo {author} {\bibfnamefont {I.}~\bibnamefont
  {Campbell}}\ and\ \bibinfo {author} {\bibfnamefont {A.}~\bibnamefont
  {Gom{\`e}s}},\ }\href
  {http://www.scopus.com/inward/record.url?eid=2-s2.0-0000478685&partnerID=40&md5=724a5f03a1300810c4a1f9a5f7df5b63}
  {\bibfield  {journal} {\bibinfo  {journal} {Proceedings of the Physical
  Society}\ }\textbf {\bibinfo {volume} {91}},\ \bibinfo {pages} {319}
  (\bibinfo {year} {1967})}\BibitemShut {NoStop}%
\bibitem [{\citenamefont {Clogston}(1962)}]{PhysRev.125.439}%
  \BibitemOpen
  \bibfield  {author} {\bibinfo {author} {\bibfnamefont {A.~M.}\ \bibnamefont
  {Clogston}},\ }\href {\doibase 10.1103/PhysRev.125.439} {\bibfield  {journal}
  {\bibinfo  {journal} {Phys. Rev.}\ }\textbf {\bibinfo {volume} {125}},\
  \bibinfo {pages} {439} (\bibinfo {year} {1962})}\BibitemShut {NoStop}%
\bibitem [{Not()}]{Note10d}%
  \BibitemOpen
  \href@noop {} {\enquote {\bibinfo {title} {Note that the total number of 10
  $d$-states needs to be conserved at the impurity site regardless of the
  number of valence electrons.}}\ }\BibitemShut {NoStop}%
\bibitem [{\citenamefont {Anderson}(1961)}]{Anderson196141}%
  \BibitemOpen
  \bibfield  {author} {\bibinfo {author} {\bibfnamefont {P.}~\bibnamefont
  {Anderson}},\ }\href
  {http://www.scopus.com/inward/record.url?eid=2-s2.0-2842617037&partnerID=40&md5=70aabfc7572d2b07ba3a27c7a9ea996c}
  {\bibfield  {journal} {\bibinfo  {journal} {Physical Review}\ }\textbf
  {\bibinfo {volume} {124}},\ \bibinfo {pages} {41} (\bibinfo {year}
  {1961})}\BibitemShut {NoStop}%
\bibitem [{\citenamefont {Friedel}(1958)}]{Friedel1958287}%
  \BibitemOpen
  \bibfield  {author} {\bibinfo {author} {\bibfnamefont {J.}~\bibnamefont
  {Friedel}},\ }\href
  {http://www.scopus.com/inward/record.url?eid=2-s2.0-51249193030&partnerID=40&md5=2014ae6fd4194d09f4433dcc70927361}
  {\bibfield  {journal} {\bibinfo  {journal} {Il Nuovo Cimento}\ }\textbf
  {\bibinfo {volume} {7}},\ \bibinfo {pages} {287} (\bibinfo {year}
  {1958})}\BibitemShut {NoStop}%
\bibitem [{\citenamefont {Mohn}(2003)}]{Mohn}%
  \BibitemOpen
  \bibfield  {author} {\bibinfo {author} {\bibfnamefont {P.}~\bibnamefont
  {Mohn}},\ }\href@noop {} {\emph {\bibinfo {title} {Magnetism in the Solid
  State An Introduction}}}\ (\bibinfo  {publisher} {Springer-Verlag},\ \bibinfo
  {year} {2003})\BibitemShut {NoStop}%
\bibitem [{\citenamefont {Andersson}\ \emph {et~al.}(2007)\citenamefont
  {Andersson}, \citenamefont {Sanyal}, \citenamefont {Eriksson}, \citenamefont
  {Nordstr\"om}, \citenamefont {Karis}, \citenamefont {Arvanitis},
  \citenamefont {Konishi}, \citenamefont {Holub-Krappe},\ and\ \citenamefont
  {Dunn}}]{PhysRevLett.99.177207}%
  \BibitemOpen
  \bibfield  {author} {\bibinfo {author} {\bibfnamefont {C.}~\bibnamefont
  {Andersson}}, \bibinfo {author} {\bibfnamefont {B.}~\bibnamefont {Sanyal}},
  \bibinfo {author} {\bibfnamefont {O.}~\bibnamefont {Eriksson}}, \bibinfo
  {author} {\bibfnamefont {L.}~\bibnamefont {Nordstr\"om}}, \bibinfo {author}
  {\bibfnamefont {O.}~\bibnamefont {Karis}}, \bibinfo {author} {\bibfnamefont
  {D.}~\bibnamefont {Arvanitis}}, \bibinfo {author} {\bibfnamefont
  {T.}~\bibnamefont {Konishi}}, \bibinfo {author} {\bibfnamefont
  {E.}~\bibnamefont {Holub-Krappe}}, \ and\ \bibinfo {author} {\bibfnamefont
  {J.~H.}\ \bibnamefont {Dunn}},\ }\href {\doibase
  10.1103/PhysRevLett.99.177207} {\bibfield  {journal} {\bibinfo  {journal}
  {Phys. Rev. Lett.}\ }\textbf {\bibinfo {volume} {99}},\ \bibinfo {pages}
  {177207} (\bibinfo {year} {2007})}\BibitemShut {NoStop}%
\bibitem [{\citenamefont {Stohr}\ and\ \citenamefont
  {Siegmann}(2006)}]{Stohr_book}%
  \BibitemOpen
  \bibfield  {author} {\bibinfo {author} {\bibfnamefont {J.}~\bibnamefont
  {Stohr}}\ and\ \bibinfo {author} {\bibfnamefont {H.~C.}\ \bibnamefont
  {Siegmann}},\ }\href@noop {} {\emph {\bibinfo {title} {Magnetism: From
  Fundamentals to Nanoscale Dynamics}}}\ (\bibinfo  {publisher} {Springer},\
  \bibinfo {year} {2006})\BibitemShut {NoStop}%
\bibitem [{\citenamefont {Schrieffer}\ and\ \citenamefont
  {Wolff}(1966)}]{PhysRev.149.491}%
  \BibitemOpen
  \bibfield  {author} {\bibinfo {author} {\bibfnamefont {J.~R.}\ \bibnamefont
  {Schrieffer}}\ and\ \bibinfo {author} {\bibfnamefont {P.~A.}\ \bibnamefont
  {Wolff}},\ }\href {\doibase 10.1103/PhysRev.149.491} {\bibfield  {journal}
  {\bibinfo  {journal} {Phys. Rev.}\ }\textbf {\bibinfo {volume} {149}},\
  \bibinfo {pages} {491} (\bibinfo {year} {1966})}\BibitemShut {NoStop}%
\bibitem [{\citenamefont {Schrieffer}(1967)}]{Schrieffer19671143}%
  \BibitemOpen
  \bibfield  {author} {\bibinfo {author} {\bibfnamefont {J.}~\bibnamefont
  {Schrieffer}},\ }\href
  {http://www.scopus.com/inward/record.url?eid=2-s2.0-0001438003&partnerID=40&md5=c01a3a9b7e26d59e5ede726646f14123}
  {\bibfield  {journal} {\bibinfo  {journal} {Journal of Applied Physics}\
  }\textbf {\bibinfo {volume} {38}},\ \bibinfo {pages} {1143} (\bibinfo {year}
  {1967})}\BibitemShut {NoStop}%
\bibitem [{\citenamefont {Yosida}(1957)}]{PhysRev.106.893}%
  \BibitemOpen
  \bibfield  {author} {\bibinfo {author} {\bibfnamefont {K.}~\bibnamefont
  {Yosida}},\ }\href {\doibase 10.1103/PhysRev.106.893} {\bibfield  {journal}
  {\bibinfo  {journal} {Phys. Rev.}\ }\textbf {\bibinfo {volume} {106}},\
  \bibinfo {pages} {893} (\bibinfo {year} {1957})}\BibitemShut {NoStop}%
\bibitem [{\citenamefont {Chien}\ \emph {et~al.}(1986)\citenamefont {Chien},
  \citenamefont {Liou}, \citenamefont {Kofalt}, \citenamefont {Yu},
  \citenamefont {Egami},\ and\ \citenamefont {McGuire}}]{PhysRevB.33.3247}%
  \BibitemOpen
  \bibfield  {author} {\bibinfo {author} {\bibfnamefont {C.~L.}\ \bibnamefont
  {Chien}}, \bibinfo {author} {\bibfnamefont {S.~H.}\ \bibnamefont {Liou}},
  \bibinfo {author} {\bibfnamefont {D.}~\bibnamefont {Kofalt}}, \bibinfo
  {author} {\bibfnamefont {W.}~\bibnamefont {Yu}}, \bibinfo {author}
  {\bibfnamefont {T.}~\bibnamefont {Egami}}, \ and\ \bibinfo {author}
  {\bibfnamefont {T.~R.}\ \bibnamefont {McGuire}},\ }\href {\doibase
  10.1103/PhysRevB.33.3247} {\bibfield  {journal} {\bibinfo  {journal} {Phys.
  Rev. B}\ }\textbf {\bibinfo {volume} {33}},\ \bibinfo {pages} {3247}
  (\bibinfo {year} {1986})}\BibitemShut {NoStop}%
\bibitem [{\citenamefont {Haag}\ \emph {et~al.}(2014)\citenamefont {Haag},
  \citenamefont {Illg},\ and\ \citenamefont {F\"ahnle}}]{PhysRevB.90.014417}%
  \BibitemOpen
  \bibfield  {author} {\bibinfo {author} {\bibfnamefont {M.}~\bibnamefont
  {Haag}}, \bibinfo {author} {\bibfnamefont {C.}~\bibnamefont {Illg}}, \ and\
  \bibinfo {author} {\bibfnamefont {M.}~\bibnamefont {F\"ahnle}},\ }\href
  {\doibase 10.1103/PhysRevB.90.014417} {\bibfield  {journal} {\bibinfo
  {journal} {Phys. Rev. B}\ }\textbf {\bibinfo {volume} {90}},\ \bibinfo
  {pages} {014417} (\bibinfo {year} {2014})}\BibitemShut {NoStop}%
\bibitem [{\citenamefont {Haag}\ \emph {et~al.}(2013)\citenamefont {Haag},
  \citenamefont {Illg},\ and\ \citenamefont {F{\"a}hnle}}]{Haag2013}%
  \BibitemOpen
  \bibfield  {author} {\bibinfo {author} {\bibfnamefont {M.}~\bibnamefont
  {Haag}}, \bibinfo {author} {\bibfnamefont {C.}~\bibnamefont {Illg}}, \ and\
  \bibinfo {author} {\bibfnamefont {M.}~\bibnamefont {F{\"a}hnle}},\ }\href
  {http://www.scopus.com/inward/record.url?eid=2-s2.0-84879720634&partnerID=40&md5=aea9f7d04c7aec1bec54e8eec24e0330}
  {\bibfield  {journal} {\bibinfo  {journal} {Physical Review B - Condensed
  Matter and Materials Physics}\ }\textbf {\bibinfo {volume} {87}} (\bibinfo
  {year} {2013})},\ \bibinfo {note} {cited By (since 1996)1}\BibitemShut
  {NoStop}%
\bibitem [{\citenamefont {Illg}\ \emph {et~al.}(2013)\citenamefont {Illg},
  \citenamefont {Haag},\ and\ \citenamefont {F\"ahnle}}]{PhysRevB.88.214404}%
  \BibitemOpen
  \bibfield  {author} {\bibinfo {author} {\bibfnamefont {C.}~\bibnamefont
  {Illg}}, \bibinfo {author} {\bibfnamefont {M.}~\bibnamefont {Haag}}, \ and\
  \bibinfo {author} {\bibfnamefont {M.}~\bibnamefont {F\"ahnle}},\ }\href
  {\doibase 10.1103/PhysRevB.88.214404} {\bibfield  {journal} {\bibinfo
  {journal} {Phys. Rev. B}\ }\textbf {\bibinfo {volume} {88}},\ \bibinfo
  {pages} {214404} (\bibinfo {year} {2013})}\BibitemShut {NoStop}%
\bibitem [{\citenamefont {Schellekens}\ and\ \citenamefont
  {Koopmans}(2013)}]{PhysRevLett.110.217204}%
  \BibitemOpen
  \bibfield  {author} {\bibinfo {author} {\bibfnamefont {A.~J.}\ \bibnamefont
  {Schellekens}}\ and\ \bibinfo {author} {\bibfnamefont {B.}~\bibnamefont
  {Koopmans}},\ }\href {\doibase 10.1103/PhysRevLett.110.217204} {\bibfield
  {journal} {\bibinfo  {journal} {Phys. Rev. Lett.}\ }\textbf {\bibinfo
  {volume} {110}},\ \bibinfo {pages} {217204} (\bibinfo {year}
  {2013})}\BibitemShut {NoStop}%
\bibitem [{\citenamefont {Kambersky}\ and\ \citenamefont
  {Patton}(1975)}]{PhysRevB.11.2668}%
  \BibitemOpen
  \bibfield  {author} {\bibinfo {author} {\bibfnamefont {V.}~\bibnamefont
  {Kambersky}}\ and\ \bibinfo {author} {\bibfnamefont {C.~E.}\ \bibnamefont
  {Patton}},\ }\href {\doibase 10.1103/PhysRevB.11.2668} {\bibfield  {journal}
  {\bibinfo  {journal} {Phys. Rev. B}\ }\textbf {\bibinfo {volume} {11}},\
  \bibinfo {pages} {2668} (\bibinfo {year} {1975})}\BibitemShut {NoStop}%
\bibitem [{\citenamefont {\ifmmode \mbox{\c{S}}\else \c{S}\fi{}a\ifmmode
  \mbox{\c{s}}\else \c{s}\fi{}\ifmmode \imath \else \i
  \fi{}o\ifmmode~\breve{g}\else \u{g}\fi{}lu}\ \emph
  {et~al.}(2011)\citenamefont {\ifmmode \mbox{\c{S}}\else \c{S}\fi{}a\ifmmode
  \mbox{\c{s}}\else \c{s}\fi{}\ifmmode \imath \else \i
  \fi{}o\ifmmode~\breve{g}\else \u{g}\fi{}lu}, \citenamefont {Friedrich},\ and\
  \citenamefont {Bl\"ugel}}]{PhysRevB.83.121101}%
  \BibitemOpen
  \bibfield  {author} {\bibinfo {author} {\bibfnamefont {E.}~\bibnamefont
  {\ifmmode \mbox{\c{S}}\else \c{S}\fi{}a\ifmmode \mbox{\c{s}}\else
  \c{s}\fi{}\ifmmode \imath \else \i \fi{}o\ifmmode~\breve{g}\else
  \u{g}\fi{}lu}}, \bibinfo {author} {\bibfnamefont {C.}~\bibnamefont
  {Friedrich}}, \ and\ \bibinfo {author} {\bibfnamefont {S.}~\bibnamefont
  {Bl\"ugel}},\ }\href {\doibase 10.1103/PhysRevB.83.121101} {\bibfield
  {journal} {\bibinfo  {journal} {Phys. Rev. B}\ }\textbf {\bibinfo {volume}
  {83}},\ \bibinfo {pages} {121101} (\bibinfo {year} {2011})}\BibitemShut
  {NoStop}%
\bibitem [{\citenamefont {Surer}\ \emph {et~al.}(2012)\citenamefont {Surer},
  \citenamefont {Troyer}, \citenamefont {Werner}, \citenamefont {Wehling},
  \citenamefont {L\"auchli}, \citenamefont {Wilhelm},\ and\ \citenamefont
  {Lichtenstein}}]{PhysRevB.85.085114}%
  \BibitemOpen
  \bibfield  {author} {\bibinfo {author} {\bibfnamefont {B.}~\bibnamefont
  {Surer}}, \bibinfo {author} {\bibfnamefont {M.}~\bibnamefont {Troyer}},
  \bibinfo {author} {\bibfnamefont {P.}~\bibnamefont {Werner}}, \bibinfo
  {author} {\bibfnamefont {T.~O.}\ \bibnamefont {Wehling}}, \bibinfo {author}
  {\bibfnamefont {A.~M.}\ \bibnamefont {L\"auchli}}, \bibinfo {author}
  {\bibfnamefont {A.}~\bibnamefont {Wilhelm}}, \ and\ \bibinfo {author}
  {\bibfnamefont {A.~I.}\ \bibnamefont {Lichtenstein}},\ }\href {\doibase
  10.1103/PhysRevB.85.085114} {\bibfield  {journal} {\bibinfo  {journal} {Phys.
  Rev. B}\ }\textbf {\bibinfo {volume} {85}},\ \bibinfo {pages} {085114}
  (\bibinfo {year} {2012})}\BibitemShut {NoStop}%
\end{thebibliography}%

\end{document}